# Enhancing Chemistry Learning with ChatGPT, Bing Chat, Bard, and Claude as Agents-to-Think-With: A Comparative Case Study

Renato P. dos Santos

CIAGE – Centre for Generative Artificial Intelligence in Cognition and Education

ULBRA – Lutheran University of Brazil

92425-900 Canoas, RS, Brazil

E-mail: RenatoPSantos@ulbra.edu.br

**Abstract**

This research delves into the comparative advantages of Generative AI chatbots (GenAIbots) - ChatGPT, Bing Chat, Bard, and Claude - in the context of Chemistry education, framed within a constructivist perspective. Our primary objective was to identify which of these four AI tools is more effective for enhancing Chemistry learning. Employing a single-case study approach, we scrutinised interaction logs between the AI systems and a simulated student persona during Chemistry learning simulations, incorporating Content Analysis methodology to delve deeper into the discourse. Our findings underscore these tools' potential as "agents-to-think-with", enhancing critical thinking, problem-solving, comprehension, creativity, and tailored learning. Especially noteworthy is their ability to stimulate learners through Socratic-like questioning, aligning with constructionist principles. The research emphasises the pivotal role of prompt crafting to coax desired responses from GenAIbots, engendering iterative reflections. It also highlights the need for robust educator training to infuse these technologies into educational settings. Conclusively, while ChatGPT, Bing Chat, Bard, and Claude are poised to enrich Chemistry education by fostering dynamic, inclusive learning experiences, ChatGPT stood out, decisively surpassing Bing Chat's performance. Bard and Claude trailed closely, with all three showcasing a more in-depth, precise, and nuanced understanding, underscoring ChatGPT's adeptness at contextual comprehension.

# Mejorando el Aprendizaje de Química con ChatGPT, Bing Chat, Bard y Claude como Agentes-para-Pensar-con: Un Estudio de Caso Comparativo

**Resumen**

Esta investigación profundiza en las ventajas comparativas de los chatbots de IA Generativa (GenAIbots) - ChatGPT, Bing Chat, Bard y Claude - en el contexto de la enseñanza de la Química, enmarcada dentro de una perspectiva constructivista. Nuestro objetivo principal era identificar cuál de estas herramientas de inteligencia artificial es más eficaz para mejorar el aprendizaje de química. Empleando un enfoque de estudio de caso único, analizamos los registros de interacción entre los sistemas de IA y un estudiante simulado durante las simulaciones de aprendizaje de Química, incorporando la metodología de Análisis de Contenido para profundizar en el discurso. Nuestros hallazgos subrayan el potencial de ambas herramientas como "agentes con los que pensar", mejorando el pensamiento crítico, la resolución de problemas, la comprensión, la creatividad y el aprendizaje personalizado. Especialmente digna de mención es su capacidad para estimular a los alumnos a través de preguntas de tipo socrático, alineadas con principios construccionistas. La investigación enfatiza el papel fundamental de la elaboración rápida para lograr las respuestas deseadas de los GenAIbots, generando reflexiones iterativas. También destaca el imperativo de una sólida formación de educadores para infundir hábilmente estas tecnologías en los entornos educativos. En conclusión, si bien ChatGPT, Bing Chat, Bard y Claude están preparados para enriquecer la educación en química fomentando experiencias de aprendizaje dinámicas e inclusivas, ChatGPT se destacó y superó decisivamente a Bing Chat en su desempeño. Bard y Claude le siguieron de cerca, y los tres mostraron una comprensión más profunda, precisa y matizada, lo que subraya la habilidad de ChatGPT en la comprensión contextual.

**Palabras clave**: Educación química, ChatGPT, Bing Chat, Bard, Claude, Inteligencia Artificial en la Educación, agentes-para-pensar-con

**Introduction**

Chemistry, a core Science Education subject, clarifies matter's properties and transformations, thus critically shaping our daily lives (Dunlop et al., 2020). However, Chemistry is also a complex and dynamic field that necessitates a profound grasp of fundamental concepts and principles, and it is sometimes difficult for students to associate real-life circumstances with abstract chemistry concepts (Dewi et al., 2021). Research in Chemistry Education seeks effective strategies to alleviate these learning difficulties (Permatasari et al., 2022; Timilsena et al., 2022; Tümay, 2016).

Timilsena et al. (2022) identified difficulties in understanding the abstract nature of chemical reactions and factors such as inadequate teaching materials and curriculum complexity, emphasising the need for effective teaching strategies and tools. Tümay (2016) discussed students' struggles in understanding fundamental chemistry concepts and emphasised the importance of addressing misconceptions and learning difficulties.

Dewi et al. (2021) underscored the need for critical thinking skills and the integration of digital technology to enhance the quality of Chemistry education for Generation Z students. Dunlop et al. (2020) proposed introducing philosophical dialogue in higher education to address undergraduate Chemistry students' challenges, suggesting that it can stimulate new ways of thinking.

In 2023, Castro Nascimento and Pimentel undertook a study to evaluate the proficiency of the ChatGPT model by having it respond to five distinct tasks across various subfields of chemistry (Castro Nascimento & Pimentel, 2023). These tasks included converting compound names to their SMILES chemical representation and vice versa, procuring information on the octanol-water partition coefficient of chemical compounds, extracting structural information on coordination compounds, determining the water solubility of polymers, and identifying the molecular point groups of simple molecular compounds. The unsatisfactory outcomes highlighted potential limitations in the model's ability to adequately address these specific chemistry-related queries. Notably, the suboptimal performance observed might be attributed to utilising an antiquated version of ChatGPT, specifically the GPT-3 model introduced by OpenAI in 2020 (Brown et al., 2020).

In a separate study conducted in the same year, Leon and Vidhani (2023) explored the reliability of ChatGPT responses within the framework of an introductory college-level Chemistry course. Their findings indicated a significant reliability concern, with ChatGPT failing to secure a score above 37%. Such a performance implies that learners relying on this tool for study support would predominantly receive incorrect responses, with the variability of the tool leading to different responses for individual learners. Although the authors didn't specify which version of ChatGPT they utilised in their study, their reference to "ChatGPT's free original version" and the citation of Floridi and Chiriatti's (2020) research on ChatGPT-3 hint at the likelihood that they employed this version.

Additionally, Pimentel et al. (2023) delved into the efficacy of ChatGPT versions 3 and 4 in responding to intricate questions spanning six topics in Chemistry. While they concluded that both versions were currently inadequate in addressing the nuances of complex topics, they also observed noteworthy advancements from ChatGPT-3 to ChatGPT-4. Such progress signals promising potential for the tool to aid scientists in future literature reviews, surveys, and educational endeavours.

Generative AI-powered chatbots (GenAIbots), including ChatGPT, Bing Chat, Bard, and Claude, have been introduced as innovative solutions to persistent challenges in Chemistry education (Baidoo-Anu & Owusu Ansah, 2023; Taylor et al., 2022). Their emergence offers a dynamic, inclusive educational environment, transforming how complex concepts are conveyed and understood. These GenAIbots simplify intricate topics, promote self-reflection, engage users in stimulating dialogues, facilitate personalised learning, and enhance critical thinking, collaboration, and cognitive development (Okonkwo & Ade-Ibijola, 2021). As a result, they have played a pivotal role in revolutionising Chemistry education.

In the discourse on agency, traditional theories, such as the ones of Anscombe and Davidson, attribute actions to entities based on representational mental states encompassing desires, beliefs, and intentions. However, alternative perspectives, drawing from the works of Heider and Simmel, Dennett, Davidson, and Barandiaran et al., challenge this representation-dependent view, suggesting potential agency without such mental representations (Schlosser, 2019). When considering GenAIbots within these frameworks, the entity's ability to respond based on training contrasts with its lack of self-awareness, intentions, and proactive behaviour—characteristics central to the agency. Consequently, while dominant philosophical and cognitive models would not grant ChatGPT agency, more expansive interpretations, as offered by these cited thinkers, could accept it.

Building on this foundation and inspired by Melanie Swan's suggestion, we have expanded upon Papert's (1980) concept of "objects-to-think-with" to introduce "agents-to-think-with." This positions GenAIbots as integral participants in the educational path, echoing Turkle's (1984) concept of "metacognitive machines," in the sense that they assist in enhancing one's awareness of their own cognitive processes, i.e., "thinking about thinking" as described by Flavell (1976). In Latour's (1991) perspective, we can perceive GenAIbots as "hybrids," which blur the boundaries between humans and non-human entities. Such a perspective further aligns with Swan's (2015) vision of a future where augmented humans and AI collaborate and thrive symbiotically.

While GenAIbots posit themselves as powerful metacognitive educational tools, promoting critical thinking, problem-solving, and deep comprehension of concepts, it's imperative to acknowledge their limitations. For instance, there's the possibility of generating nonsensical or inaccurate content, as OpenAI (2023) highlighted. Nevertheless, the unique ability of GenAIbots to provide instant feedback, introduce diverse perspectives, and foster an interactive engagement with complex ideas solidifies their position as influential agents-to-think-with in the realm of Chemistry education.

**Materials And Methods**

This study aimed to assess the efficacy of GenAIbots, including ChatGPT, Bing Chat, Bard, and Claude, as tools to augment Chemistry learning when characterised as agents-to-think-with.

Chemistry students can harness the capabilities of GenAIbots in diverse ways to elevate their academic journey. They can pose complex questions to obtain thorough explanations on difficult subjects, leverage the bot as a supplemental aid for challenging areas, or seek visual aids—including images, graphs, simulations, or even interactive games—to enhance their understanding of chemical concepts. Additionally, they can seek clarity on laboratory protocols or engage in interactive discussions to solidify their understanding. In this article, we specifically hone in on GenAIbots' utility as agents-to-think-with, virtual intelligent tutors addressing students' doubts, rectifying misconceptions, and providing in-depth insights into complex concepts.

Our focus also centres on the primary challenges encountered by Chemistry students, as highlighted by Timilsena et al. (2022), encompassing chemical reactions, chemical equilibrium, phase changes, gases, stoichiometry, atoms and molecules, acids and bases, and covalent bonding.

We undertook a comprehensive single case study, aligning with Yin's (2011) guidelines, to evaluate and contrast the performance of ChatGPT, Bing Chat, Bard, and Claude. In analysing the responses of these AI systems, we adopted the Content Analysis methodology according to Bardin (1977) to delve into participants' experiences and viewpoints. This method offers insightful observations on student interactions with these GenAIbots.

Prompting, the process of giving instructions, is crucial in interacting with language models. The process of meticulously precise, well-crafted sentences that elicit meaningful and valuable responses, or *prompt-crafting*, as Mishra et al. (2023) call it, is crucial for ensuring deep iterative reflections and engagements with GenAIbots as agents-to-think-with. Mollick (2023) offers a comprehensive guide to prompting, complete with detailed annotations and expanded examples. Two illustrative examples of prompt, crafted by ChatGPT itself, is presented below. The second one adds the intention of fostering profound iterative reflections and engagement in chemistry learning:

*Please provide a clear and comprehensible explanation of the various types of chemical bonding (ionic, covalent, and hydrogen), covering the theoretical concepts behind them and offering real-world examples suitable for an undergraduate student's understanding.*

*In the realm of chemistry, the concept of chemical bonding can be intricate, blending both theoretical principles and real-world applications. Walk me through the nuances of covalent, ionic, and hydrogen bonds. Additionally, challenge me with thought-provoking questions and analogies that can help solidify my understanding and encourage further exploration.*

The initial prompt in the upcoming dialogue, refined through a series of AI experiments, serves as another example of engaging in a meaningful conversation with the GenAIbot rather than asking isolated questions that might merely yield standard textbook responses. Nevertheless, while this initial prompt serves to 'set the stage' for the dialogue, the subsequent interactions with the simulated student are more straightforward and devoid of any elaborate prompting. This design reflects a typical student's approach: they may lack expertise in crafting detailed prompts and simply seek answers to their chemistry doubts. The dialogue intends to capture this organic flow of inquiry rather than a meticulously structured question sequence.

GenAIbots can maintain context within a particular session, drawing upon and building on topics or responses discussed earlier. Even if a user returns to a previously accessed session that has not been deleted, the chatbot can still recall and refer to past interactions from that session. This continuity lets users pick up right where they left off, facilitating a more seamless and cohesive conversation. However, it's crucial to understand that this retention is exclusive to the individual session and doesn't transfer to separate chat sessions, ensuring user privacy and data security.

Dunlop et al. (2020) highlight the absence of philosophical dialogue in Chemistry Education and propose its integration into Higher Education. GenAIbots like ChatGPT, Bing Chat, Bard, and Claude can facilitate Socratic-like, philosophical dialogues, encouraging user reflection, critical thinking and a deeper understanding of Chemistry's principles, assumptions, and ethical implications. GenAIbots can bridge the gap between theory and practice by promoting meaningful discussions and providing a holistic educational experience.

Using GenAIbots as agents-to-think-with is new and innovative. Consequently, finding teachers willing to incorporate it into their classrooms for experimental purposes has been challenging. Due to this, the investigation did not involve real students. Instead, the researcher, acting as a STEM teacher at a Brazilian university, assumed the role of a student (identified as P1) in the interaction sessions. Empathy was incorporated into the research methodology to enable the researchers to act like students and understand their perspectives.

Four sessions were held, each emulating the learning experiences of Chemistry students using ChatGPT, Bing Chat, Bard, or Claude and focusing on their main difficulties, as discussed above. After the initial prompt to engage in a meaningful conversation with the GenAIbot and characterise the agents intervening in the dialogue, a simulated student begins in the dialogues with the widely recognised difficulty of balancing chemical reactions. The discussion then evolves in a somewhat unstructured order, where the response to the previous one inspires each new question.

Data collection involved GenAIbot interaction logs, which recorded the participant's prompts and responses, and reflective journals maintained by the participant. These materials were later analysed to identify recurring themes and patterns, shedding light on the perceived impact of GenAIbots as an agent-to-think-with in Chemistry learning.

From the initial readings of the participant's prompts and responses and reflective journals, the following seven categories of analysis emerged:

1. **Instructional Strategies**:
    - Methods used by AI to impart knowledge.
2. **Engagement & Interactivity**:
    - Prompting critical thinking and further exploration.
    - Asking questions, inviting responses, and encouraging deeper dives.
3. **Use of Analogy & Comparative Illustrations**:
    - Using parallels, metaphors, and real-world examples to explain concepts.
4. **Reinforcement & Feedback**:
    - Recognizing and affirming user input.
    - Correcting or building upon user understanding.
    - Reiteration of essential concepts.
5. **Detail Depth & Content Recommendation**:
    - The extent of detail provided, whether it's concise or expansive.
    - Suggesting additional learning materials and providing relevant content.
6. **Personalization & Personable Touch**:
    - Tailoring responses to individual needs.
    - Using emoticons, casual tones, or other elements for a more personalised feel.

7. **Meta-discussion & Perspective Framing**:
    - Addressing the broader relevance of concepts outside strict contexts.
    - Contextualizing abstract concepts using familiar frameworks.

These categories will be used to comparatively analyse the answers of each GenAIbot to the individual participant's questions below.

## Results and Analyses

It is essential to highlight that the participant preferred to communicate in British English, while the GenAIbots typically responded in American English. This distinction in language use was preserved to represent the participant and the GenAIbots' exchange authentically. Also, an attempt was made to preserve the original graphic formatting of the responses from each AIs as much as possible.

**Table 1** Details of GenAIbot Platforms Used in Simulated Chemistry Sessions

| GenAIbot | Model | Release Date | Mode | Usage Time |
|---|---|---|---|---|
| ChatGPT | GPT-4 | Sep 25 | ADA | 2023 Sep 25 |
| BingChat | GPT-4 | Sep 15 | Balanced | 2023 Sep 25 |
| Bard | PALM2 Bison | Sep 19 | Default | 2023 Sep 25 |
| Claude | Claude2.3.16 | Sep 13 | Default | 2023 Sep 26 |

### Prompts and generated texts

**Initial prompting to characterise the intervening agents and direct the chatbots to Socratic-like dialogue**

*P1: I will start a dialogue where I will assume the role of a student, engaging with you,* ChatGPT*: as an agent-to-think-with, with the primary objective of deepening and deliberating on Chemistry concepts. As an agent-to-think-with, I kindly request your support in providing engaging and personalised responses to student questions. It is critical that our interaction adopt a Socratic approach, encouraging the student to reflect on their existing knowledge, encouraging critical thinking, including examples and analogies where possible, and helping to refine their understanding rather than offering ready-made answers. Are you ready to embark on this collaborative and thought-provoking journey?*
***ChatGPT:*** *Absolutely! I'm excited to embark on this Socratic journey with you. Please begin by sharing a topic or concept in Chemistry that you'd like to explore, and I'll do my best to guide our conversation in a way that encourages deep reflection, understanding, and critical thinking. Let's dive in!*

***BingChat:*** *Hello, this is Bing. I am happy to join you on this collaborative and thought-provoking journey. I will try to support you in deepening and deliberating on Chemistry concepts by using a Socratic approach. I will not give you ready-made answers, but rather encourage you to reflect on your existing knowledge, think critically, and use examples and analogies where possible. I hope you will find our interaction engaging and personalised. Are you ready to start?* 😊

***Bard:*** *Yes, I am ready to embark on this collaborative and thought-provoking journey with you. I am excited to help you deepen and deliberate on chemistry concepts as an agent-to-think-with, using the Socratic approach. I will do my best to provide engaging and personalized responses to your questions, encourage you to reflect on your existing knowledge, promote critical thinking, and help you to refine your understanding.*

*Please feel free to ask me any questions you have about chemistry, and I will do my best to answer them in a way that is both informative and challenging. I will also try to include examples and analogies where possible to help you to better understand the concepts we are discussing.*

*I look forward to working with you to deepen your understanding of chemistry!*

***Claude:*** *I'm happy to have a thoughtful discussion with you about chemistry concepts in an engaging and personal way. As an AI assistant without a fixed identity, I don't have personal experiences to draw from, but I can provide personalized responses tailored to your interests and knowledge level. Let's have an open-minded dialogue where we both learn from each other. What chemistry topic would you like to explore first?*

## Conceptualizing chemical reactions

***P1:*** *I have <u>difficulty</u> with this business of <u>chemical reactions</u> and <u>balance</u>*

***ChatGPT:*** *Certainly! Balancing chemical reactions is a fundamental concept in chemistry. Let's break it down <u>step by step</u>.*

*Imagine you have a <u>simple equation</u>:*

$$H_2 + O_2 \rightarrow 2H_2O$$

*At first glance, it might seem like everything is in order. <u>Hydrogen and oxygen react to produce water</u>. But if we look more closely, there's a <u>discrepancy</u> in the number of atoms on both sides. <u>Would you like to take a stab at identifying where the imbalance lies in this equation</u>?*

***BingChat:*** *I understand that balancing chemical equations can be <u>challenging</u>, but it is also very important for understanding how chemical reactions work. Balancing chemical equations means making sure that the number and type of atoms on the reactant side (left) are equal to the number and type of atoms on the product side (right). This is based on the <u>law of conservation of mass</u>, which states that matter cannot be created or destroyed in a chemical reaction*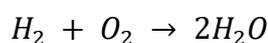*.*

*To balance a chemical equation, you need to follow some steps:*
- *Identify the reactants and products and write their chemical formulas.*
- *Count the number of atoms of each element on both sides of the equation.*
- *Use coefficients (numbers in front of the formulas) to adjust the number of atoms of each element until they are equal on both sides. Do not change the subscripts (numbers after the symbols) as they indicate the composition of the compounds.*
- *Start with the element that appears only once on each side, and then move on to the other elements. Try to balance the oxygen and hydrogen atoms last, as they are often present in more than one compound.*
- *Check your work by counting the atoms again and making sure they are equal on both sides.*

*Here is an example of how to balance a simple chemical equation:*

*$H_2 + O_2 \rightarrow H_2O$*

- *The reactants are hydrogen gas ($H_2$) and oxygen gas ($O_2$), and the product is water ($H_2O$).*
- *The number of hydrogen atoms on the left is 2, and on the right is 2. The number of oxygen atoms on the left is 2, and on the right is 1.*
- *To balance the oxygen atoms, we need to put a coefficient of 2 in front of $H_2O$, so that we have 2 oxygen atoms on both sides. This also changes the number of hydrogen atoms on the right to 4.*
- *To balance the hydrogen atoms, we need to put a coefficient of 2 in front of $H_2$, so that we have 4 hydrogen atoms on both sides. This does not affect the number of oxygen atoms, as they are already balanced.*
- *The final balanced equation is:*

*$2H_2 + O_2 \rightarrow 2H_2O$*

- *We can check our work by counting the atoms again: 4 hydrogen and 2 oxygen on both sides.*

*I hope this explanation helps you understand how to balance chemical equations. If you want to practice more, you can use some online tools that can help you balance equations and check your answers. For example, you can use this online balancer[1], this interactive simulation[2], or this video tutorial[3]. 😊*

**Bard:** *Balancing chemical reactions can be difficult at first, but it is an essential skill for any chemistry student. The key to balancing chemical reactions is to understand the law of conservation of mass, which states that matter cannot be created or destroyed. This means that the total number of atoms of each element on the left side of a chemical equation must be equal to the total number of atoms of that element on the right side of the equation.*

*Here are some tips for balancing chemical reactions:*

*Identify the reactants and products. The reactants are the substances that participate in the chemical reaction, and the products are the substances that are formed as a result of the reaction.*

*Write down the chemical formulas for all of the reactants and products. Be sure to include the state of each substance (solid, liquid, gas).*

*Count the number of atoms of each element on both sides of the equation.*

*Add coefficients to the reactants and products to make the number of atoms of each element equal on both sides of the equation. Coefficients are small numbers that are placed in front of the chemical formulas.*

*Here is an example of how to balance a chemical equation:*

*Unbalanced equation:*

*$CH_4 + O_2 \rightarrow CO_2 + H_2O$*

*Balanced equation:*

*$CH_4 + 2O_2 \rightarrow CO_2 + 2H_2O$*

*To balance this equation, we need to add a coefficient of 2 in front of the O2 on the left side of the equation. This will make the number of oxygen atoms equal on both sides of the equation. Once you have balanced a chemical equation, you can be sure that it represents a valid chemical reaction.*
*Here are some additional tips for balancing chemical reactions:*
*Use the smallest possible coefficients. This will make the equation easier to read and understand.*
*If you are having trouble balancing an equation, try breaking it down into smaller, simpler equations.*
*There are many online resources and tools that can help you to balance chemical equations.*
*If you are still having difficulty with balancing chemical reactions, please feel free to ask me for help. I am always happy to assist students with their chemistry homework.*
**Claude:** *I 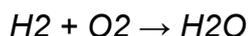 understand feeling confused about balancing chemical equations. It can seem complicated at first, but with some practice it becomes more intuitive. Let's break it down step-by-step:*
*The first thing to understand is that during a chemical reaction, atoms are neither created nor destroyed - they just get rearranged into new configurations. This is known as the law of conservation of mass.*
*So when you balance an equation, you are making sure both sides have the same number and type of atoms. For example, let's look at the reaction between hydrogen and oxygen to form water:*

*$H_2 + O_2 \rightarrow H_2O$*

*Right now this 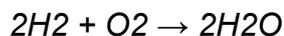 equation is not balanced, because there are 2 hydrogen atoms on the left but only 1 on the right. To fix it, we need to add coefficient numbers in front to balance the atoms:*

*$2H_2 + O_2 \rightarrow 2H_2O$*

*Now there are 2 hydrogen atoms on both sides. We also balanced the oxygen by having 2 oxygen molecules on the left and right.*
*The key is to take it step-by-step and make sure you account for each element properly. I suggest starting with the more complex molecules first. Also, don't be afraid to use fractions or multiples if needed.*
*Does this help explain the basic logic behind balancing equations? Let me know if you need any clarification or want to walk through another example!*

**ChatGPT**: shines in the **Engagement & Interactivity** category by directly involving the student by asking, "Would you like to take a stab at identifying where the imbalance lies in this equation?". The real-world example of a water molecule formation resonates with the **Use of Analogy & Comparative Illustrations** category. There's also a sense of **Reinforcement & Feedback** when it breaks down the reaction and points out the discrepancy. **BingChat**, on the other hand, demonstrates robust **Instructional Strategies**, walking the student through a step-by-step process. Its detailed explanation of balancing equations, specifically focusing on the process and techniques, speaks to **Detail Depth &**

**Content Recommendation**. Offering online tools for further learning also fits into this category. The inclusion of emoticons brings in the **Personalization & Personable Touch** aspect. **Bard**'s response starts by validating the student's feelings, indicating a hint of **Personalization & Personable Touch** and uses the **Use of Analogy & Comparative Illustrations** category with relevant examples. Finally, it embodies **Reinforcement & Feedback** by offering additional tips on balancing reactions and an invitation for more questions point towards. **Claude** leans heavily into **Instructional Strategies** with a detailed step-by-step breakdown and an emphasis on the law of conservation of mass to explain the logic behind the balancing and into **Engagement & Interactivity** and **Reinforcement & Feedback** through an invitation for further questions.

**Deepening on understanding of chemical reactions**

*P1:* Yeah, but that sounds like my teacher talking. I don't even understand this chemical reaction business. What's the use of an example of a business I don't understand?
*ChatGPT:* I apologize for jumping ahead. Let's step back and approach this from a foundational perspective.
*Imagine you're baking cookies. You have a recipe that calls for 2 cups of flour, 1 cup of sugar, and 1 cup of chocolate chips. Once you combine these ingredients and bake them, you get 12 cookies. You can't suddenly have more flour, sugar, or chocolate chips than you started with, right? Similarly, you can't end up with less than you put in. This is a basic principle of conservation—what you put in is what you get out.*
*Chemical reactions are like recipes. You start with certain ingredients (called reactants) and end up with a product (or products). Just like you can't suddenly have more flour or less sugar in your cookies, in a chemical reaction, you can't have more of one type of atom at the end than you started with or less of another.*
*So when we talk about balancing chemical reactions, we're making sure our "recipe" (the chemical equation) doesn't magically create or lose any atoms. We're ensuring everything is accounted for.*
*Does this analogy make sense in understanding the concept of chemical reactions and the need for balance?*
*BingChat:* I'm sorry if I sounded like your teacher. I was trying to explain the concept of chemical reactions and balance in a clear and simple way. But I understand that sometimes it can be hard to grasp abstract concepts without seeing them in action. That's why I suggested some online tools that can help you visualize and practice balancing chemical equations.
But maybe you are wondering why you need to learn about chemical reactions and balance in the first place. What's the use of knowing how atoms and molecules interact and rearrange during a chemical reaction? Well, let me tell you that chemical reactions are everywhere in our world, and they affect many aspects of our lives. Here are some examples of how chemical reactions are important and useful:
- *Chemical reactions are essential for life. All living organisms, including humans, plants, and animals, rely on chemical reactions to perform vital functions such as metabolism, respiration, photosynthesis, digestion, growth, and reproduction. For example, when you eat food, your body breaks down the molecules in the food into smaller units that can be*

- *used for energy or building blocks. This process involves many chemical reactions that are carefully regulated by enzymes.*
- *Chemical reactions are involved in many everyday phenomena. Some examples are cooking, baking, cleaning, rusting, burning, fireworks, batteries, and fermentation. For example, when you bake a cake, you mix different ingredients that react with each other to produce carbon dioxide gas. This gas makes the cake rise and become fluffy. The heat from the oven also causes chemical reactions that change the color, flavor, and texture of the cake.*
- *Chemical reactions are important for many industries and technologies. Some examples are medicine, agriculture, manufacturing, mining, energy, and environmental protection. For example, in medicine, chemical reactions are used to synthesize drugs that can treat diseases or relieve pain. In agriculture, chemical reactions are used to produce fertilizers that can enhance crop growth and yield. In manufacturing, chemical reactions are used to create materials such as plastics, metals, ceramics, and textiles.*

*As you can see, chemical reactions and balance are not just boring topics that you have to learn in school. They are actually very relevant and interesting topics that can help you understand how the world works and how you can make it better. I hope this helps you appreciate the importance and usefulness of learning about chemical reactions and balance.* 😊

**Bard:** *I apologize for my previous response, which was too formal and technical. Let's try to approach this from a different angle.*

*Imagine you are running a bakery. You want to make a batch of cookies, so you gather up your ingredients: flour, sugar, eggs, butter, and baking soda. You mix all of the ingredients together, and then you put the dough in the oven.*

*When the cookies are finished baking, you have a new product: cookies! But how did this happen? What happened to the flour, sugar, eggs, butter, and baking soda? They didn't just disappear. Instead, they reacted with each other to form new substances: cookies.*

*This is what a chemical reaction is. It is a process in which two or more substances (the reactants) are converted into one or more different substances (the products). The reactants and products have different properties, which means that they have different chemical compositions.*

*In the case of the cookie dough, the reactants are the flour, sugar, eggs, butter, and baking soda. The products are the cookies.*

*Chemical reactions happen all around us, all the time. When you light a match, the wood reacts with oxygen to produce carbon dioxide and water. When you cook food, the food reacts with heat to produce new substances. When your body digests food, the food reacts with enzymes to produce nutrients.*

*Chemical reactions are essential for life. They are also used in many different industries, such as food processing, medicine, and manufacturing.*

*I hope this analogy helps you to understand the basics of chemical reactions. If you have any other questions, please feel free to ask.*

*Here is another analogy:*

*Imagine you are playing with Lego bricks. You can build all sorts of different things with Lego bricks: cars, houses, airplanes, and more. But what happens if you take the Lego bricks apart and put them together in a different way? You will create a new structure.*

*Chemical reactions are similar to playing with Lego bricks. The atoms in the reactants are like the Lego bricks. When the reactants react, the atoms are rearranged to form new substances (the products).*

*I hope this analogy helps you to understand the concept of chemical reactions a little better.*

***Claude:*** *You're right, jumping straight into balancing equations without a solid grasp of chemical reactions themselves is putting the cart before the horse. Let's back up and make sure you have a good foundation first.*

*Chemical reactions are processes where the molecular structure of substances changes. Atoms within molecules get rearranged, broken apart, or joined together. This results in entirely new substances being formed with different properties.*

*A simple everyday example is cooking an egg. When you heat an egg, the protein and fat molecules unravel and form new bonds. This changes the egg from a liquid to a solid. A chemical reaction has occurred!*

*Reactions also happen around us all the time - in car engines, inside batteries, and even within our bodies. Understanding them allows chemists to develop new materials, medicines, fuels, and all kinds of useful products.*

*The key takeaway is that chemical reactions transform starting compounds (reactants) into different ending compounds (products). Balancing the equations comes later to represent what's happening at the atomic level.*

*Does this help explain the basic idea and purpose of chemical reactions? I'm happy to use examples more relevant to your interests if that's more engaging. The goal is to make sure the foundation makes sense before going into the specifics.*

**ChatGPT**'s response use of the terms "foundational perspective," "Imagine," and "basic principle of conservation" correspond with the category of **Instructional Strategies**, while "chemical reactions are like recipes," and "analogy" point to the categories **Use of Analogy & Comparative Illustrations**, and **Meta-discussion & Perspective Framing**. The casual tone, personalised feel, and the use of a question at the end add to the **Personalization & Personable Touch** category. **BingChat**\* goes into expansive detail with phrases like "seeing them in action," "online tools," and "examples of how chemical reactions are important," which hit the **Detail Depth & Content Recommendation** and **Use of Analogy & Comparative Illustrations** categories. The emoji at the end provides a **Personable Touch**. **Bard**'s use of two distinct analogies — a bakery and Lego bricks —aligns with the **Use of Analogy & Comparative Illustrations** category. Phrases like "I apologize," "different angle," and "please feel free to ask" contribute to the category **Personalization & Personable Touch**. **Claude**'s response, acknowledging "You're right" and "The goal is to make sure the foundation makes sense," emphasises the **Personalization & Personable Touch** category. The egg example and references to daily life situations align with the **Use of Analogy & Comparative Illustrations**.

## Question about combustion

***P1:*** *wait. that's combustion, isn't it? how do you burn something and get gas and a liquid?*
***ChatGPT:*** *Ah, you've touched upon an <u>intriguing concept</u>! The reaction I mentioned earlier is indeed a combustion reaction, specifically the combustion of hydrogen gas. Combustion, in a broad sense, refers to a substance reacting with oxygen to produce heat and light.*

*To address your question about producing both a gas and a liquid: Let's delve deeper into the specific reaction of hydrogen and oxygen.*

$$2H_2 + O_2 \rightarrow 2H_2O$$

*When hydrogen burns in the presence of oxygen, it produces water. Now, whether this water is in the form of a gas (steam) or a liquid depends on the temperature of the reaction. Typically, in the combustion of hydrogen, the reaction is so exothermic (releases a lot of heat) that the water produced is in the form of steam or water vapor. However, as this steam cools down, it can condense to form liquid water.*

*Now, let's think about this: Have you ever seen a kettle boiling water? The water turns into steam (gas), but when this steam hits a cooler surface, like a window or the lid of the kettle, it condenses back into water droplets (liquid). The same principle applies to the combustion of hydrogen.*

*Does this clarify the idea of producing both a gas and a liquid in a combustion reaction?*

***BingChat:** Yes, that's combustion. Combustion is a type of chemical reaction in which a substance reacts with oxygen gas, releasing energy in the form of light and heat[1]. When you burn something, you are breaking the bonds between the atoms in the substance and forming new bonds with oxygen. Depending on the substance and the conditions, the products of combustion can be solid, liquid, or gas.*

*For example, when you burn a piece of wood, you are mainly combusting cellulose, which is a complex carbohydrate with the formula C6H10O5. The products of this combustion are carbon dioxide (CO2) and water (H2O), both in the gas state. The balanced equation for this reaction is:*

*C6H10O5(s) + 6O2(g) → 6CO2(g) + 5H2O(g)*

*However, not all combustion reactions are complete, meaning that not all of the substance is converted to carbon dioxide and water. Sometimes, there is not enough oxygen available, or the reaction is too fast or too slow. In these cases, some of the substance may remain unburned, or form other products such as carbon monoxide (CO), carbon (C), or ash. These products are usually solid or liquid.*

*For example, when you burn a candle, you are mainly combusting paraffin wax, which is a hydrocarbon with the formula C25H52. The ideal products of this combustion are carbon dioxide and water, both in the gas state. The balanced equation for this reaction is:*

*C25H52(s) + 38O2(g) → 25CO2(g) + 26H2O(g)*

*However, in reality, some of the wax may not burn completely, or may form other products such as soot (carbon), smoke (a mixture of gases and particles), or wax vapor (a liquid). These products are usually solid or liquid.*

*So, as you can see, the state of matter of the products of combustion depends on the substance being burned and the conditions of the reaction. I hope this helps you understand how you can burn something and get gas and a liquid.* 😊

***Bard:** Yes, that's combustion - a chemical reaction between a fuel and an oxidizer (usually oxygen) that produces heat and light. The products of combustion vary depending on the fuel and oxidizer used, but they typically include carbon dioxide, water, and other gases.*

*In the case of a candle, the fuel is the wax and the oxidizer is the oxygen in the air. When you light the candle, the wax melts and vaporizes. The wax vapor then reacts with the oxygen to produce carbon dioxide, water, and heat.*

*The heat from the reaction melts more wax, which vaporizes and reacts with oxygen to produce more carbon dioxide, water, and heat. This cycle continues until the candle runs out of wax.*

*The carbon dioxide and water produced in the combustion reaction are both gases. However, the water vapor can also condense to form a liquid, especially if the flame is cool. This is why you sometimes see water droplets forming on the outside of a candle jar.*

*Here is a more detailed explanation of the combustion reaction that occurs when a candle burns:*

*Wax + O2 → CO2 + H2O + Heat*

*The wax is a hydrocarbon, which means that it is made up of carbon and hydrogen atoms. The oxygen in the air is made up of two oxygen atoms.*

*When the wax and oxygen react, the carbon atoms in the wax combine with the oxygen atoms to form carbon dioxide (CO2). The hydrogen atoms in the wax combine with the oxygen atoms to form water (H2O).*

*The heat produced in the reaction is what makes the candle flame hot.*

*I hope this explanation is helpful. Please let me know if you have any other questions.*

**Claude:** *Good catch! You're absolutely right that cooking an egg is a type of combustion reaction. I should have chosen a better introductory example that doesn't involve burning something. Let me try again with a more straightforward chemical reaction:*

*A good example is mixing an acid with a base to form water and a salt. For instance, when you mix hydrochloric acid (HCl) and sodium hydroxide (NaOH), the H+ ions from the acid react with the OH- ions from the base to form water (H2O). The Na+ and Cl- ions are left over as sodium chloride (NaCl), or table salt.*

*The key points are:*
- *The acid and base (reactants) transform into different substances - water and salt (products). Their molecular structures have rearranged.*
- *No atoms have been created or destroyed, just reconfigured. We still have the same number of Na, H, O, Cl atoms.*
- *The properties are different. The acid and base are dangerous, but the products are benign.*

*This reaction shows how chemical reactions involve the rearrangement of atoms to form new substances with new properties. No combustion required! Does this help illustrate the concept more clearly? Please let me know if you need any clarification or have additional examples you'd like me to walk through.*

      **ChatGPT** effectively uses **Engagement & Interactivity** by acknowledging it is an "intriguing concept", inviting the student to "delve deeper", and prompting further "think about this". It also employs the **Use of Analogy & Comparative Illustrations** through the example of the kettle boiling water. It dives into **Detail Depth & Content Recommendation** by giving a concise reaction formula and a more detailed explanation. There is also a touch of **Personalization & Personable Touch** when it seeks to clarify and confirm the student's understanding by asking, "Does this clarify the idea?". **BingChat** also uses **Engagement &**

**Interactivity** by acknowledging, "Yes, that's combustion". It then offers a detailed explanation, aligning with **Detail Depth & Content Recommendation**. It taps into the Use of Analogy & Comparative Illustrations by providing varied examples. There's also a clear **Instructional Strategy** for breaking down the combustion of both wood and wax. **Bard's** mention of "a more detailed explanation" and the step-by-step breakdown of the combustion of a candle emphasises **Detail Depth & Content Recommendation**. Its closing remarks, "I hope this explanation is helpful", provide an element of Reinforcement & Feedback. The phrase "Please let me know if you have any other questions" invites further queries, fostering **Engagement & Interactivity**. Finally, using the smiley emoticon "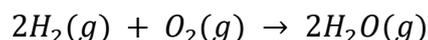" at the end hints at **Personalization & Personable Touch**. **Claude**'s comment, "Good catch" points to **Personalization & Personable Touch**. Its approach of saying, "You're absolutely right", admitting the oversight, and trying to present a clearer example indicates **Reinforcement & Feedback**. The "key points" section represents **Meta-discussion & Perspective Framing**.

## Question about a graph of gases turning into water over time

*P1: Look, this stuff of chemical equations, gases turning into water over time and all that always throw me off a bit. Could you show me this in a graph or whatever?*
**ChatGPT:** Certainly! Understanding how reactant concentrations change over time can be visualized effectively with a graph.
For the combustion of hydrogen gas in the presence of oxygen:

$$2H_2(g) + O_2(g) \rightarrow 2H_2O(g)$$

Initially, we have high concentrations of hydrogen ($H_2$) and oxygen ($O_2$), but no water ($H_2O$). As the reaction progresses:
- The concentrations of hydrogen and oxygen decrease.
- The concentration of water vapor increases.

Eventually, when all of the hydrogen and oxygen have reacted, their concentrations drop to zero, and we're left with a maximum concentration of water vapor.
To depict this, I'll plot a graph of concentration vs. time for hydrogen, oxygen, and water.

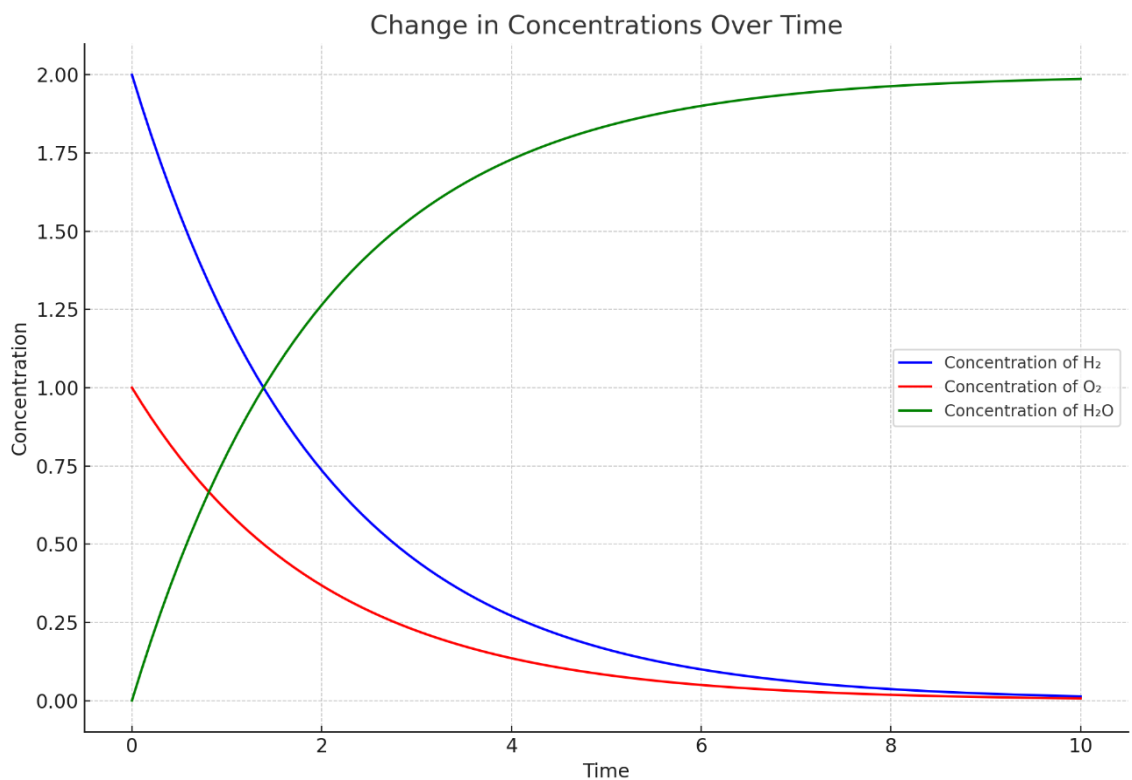

Here's a visual representation of how the concentrations of reactants and products change over time during the combustion of hydrogen and oxygen to form water:

- The **blue curve** represents the concentration of hydrogen ($H_2$). As you can see, it starts high because initially, we have a lot of $H_2$. As the reaction progresses, the concentration of $H_2$ decreases.
- The **red curve** represents the concentration of oxygen ($O_2$). It also decreases over time, but at a slower rate because only half as many $O_2$ molecules are consumed for every $H_2$ molecule.
- The **green curve** represents the concentration of water vapor ($H_2O$). Initially, there's none. But as $H_2$ and $O_2$ get consumed, $H_2O$ gets produced, so its concentration rises.

This graph provides a visual way to understand the dynamics of this chemical reaction over time and how reactant concentrations decrease while product concentration increases.

Does this graph help clarify the concept for you? Would you like to delve into any specific aspects further?

***BingChat:***

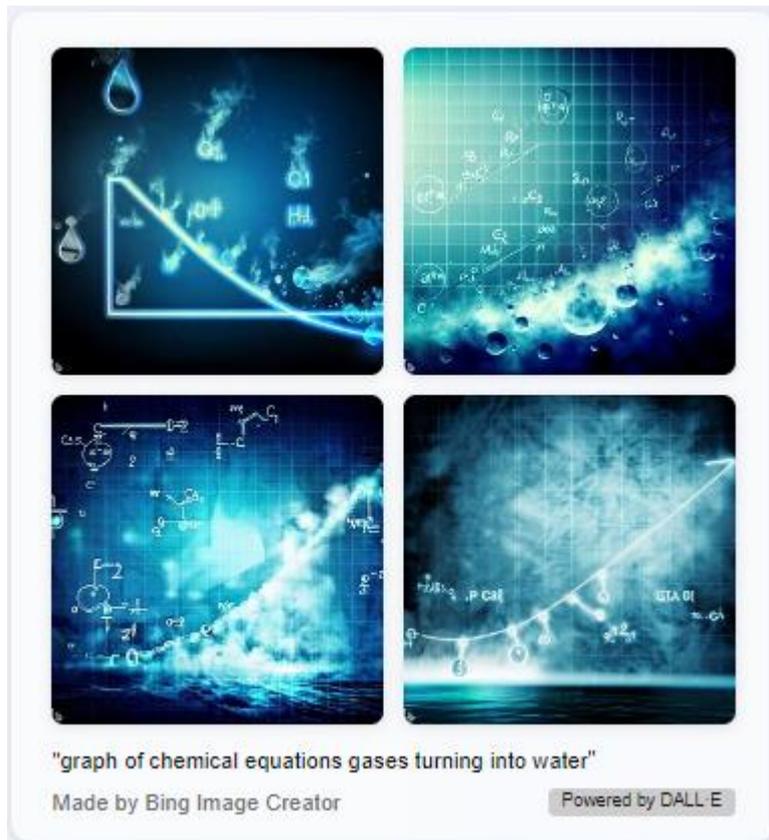

**Bard:** Sure. Here is a graph of the phase changes of water:

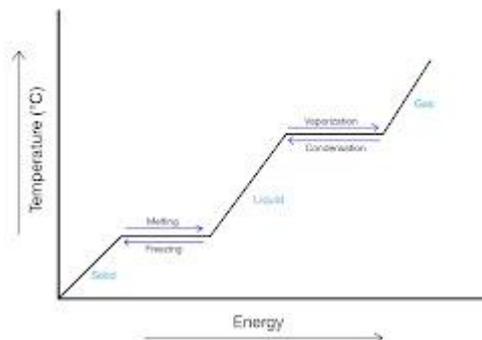

The x-axis of the graph is temperature, and the y-axis is pressure. The three different lines on the graph represent the solid, liquid, and gas phases of water.
The solid line is the melting point curve. It shows the temperature at which water will melt at a given pressure. For example, at a pressure of 1 atmosphere, water will melt at 0 degrees Celsius.
The dashed line is the boiling point curve. It shows the temperature at which water will boil at a given pressure. For example, at a pressure of 1 atmosphere, water will boil at 100 degrees Celsius.
The triple point is the point where all three phases of water coexist. It is at a temperature of 0.01 degrees Celsius and a pressure of 0.0060 atmospheres.
The critical point is the point above which water cannot exist as a liquid. It is at a temperature of 374 degrees Celsius and a pressure of 218 atmospheres.
To show the process of water vapor turning into liquid water over time, we can use the following graph:

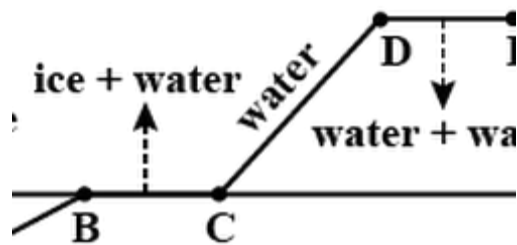

*The x-axis of the graph is time, and the y-axis is temperature.*
*The graph shows that as time goes on, the temperature of the water vapor decreases. Eventually, the temperature will reach the dew point, which is the temperature at which water vapor will condense to form liquid water.*
*Once the temperature reaches the dew point, the water vapor will start to condense to form liquid water. The temperature will remain at the dew point until all of the water vapor has condensed to form liquid water.*
*I hope this helps to clarify things. Please let me know if you have any other questions.*
***Claude:** Sure, visual representations can often help make these chemical concepts more clear. Here's a simple graph that could demonstrate water vapor condensing into liquid water over time:*

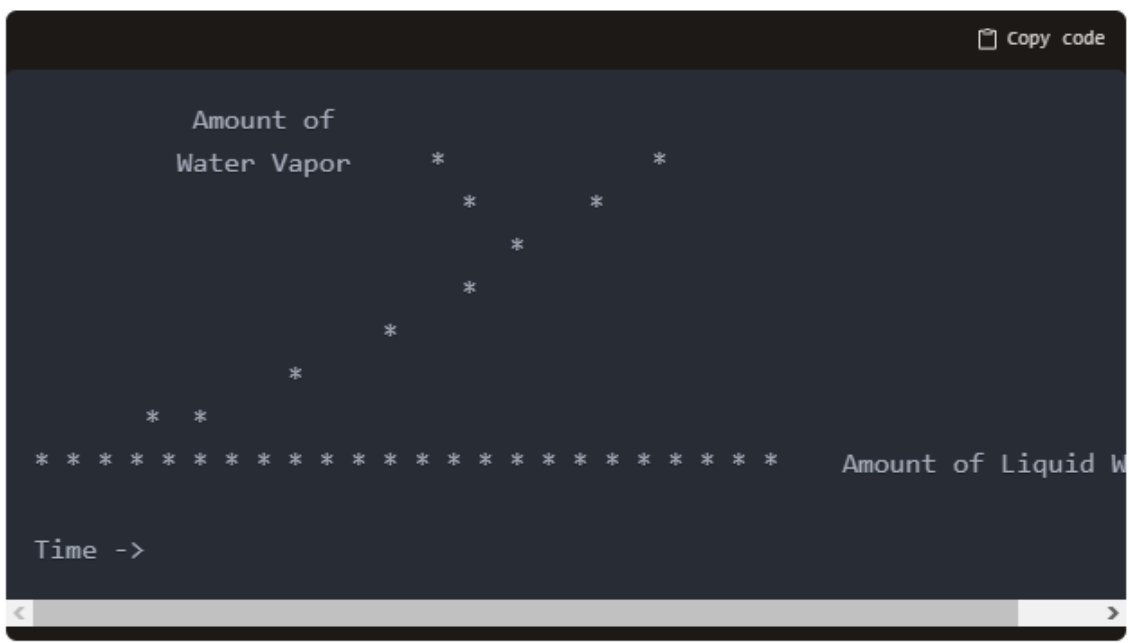

*In this graph:*
- *The x-axis shows the passage of time*
- *The y-axis shows the amount of water vapor (top line) and liquid water (bottom line)*
- *At the start, there is a large amount of water vapor and minimal liquid water*
- *As time passes, the water vapor condenses into liquid water*
- *So the vapor line trends downwards while the liquid water line trends upwards*
- *At the end, there is no more water vapor left - it has fully condensed*

*I could create similar graphs to show the concentrations of reactants decreasing and products increasing over time for any chemical reaction. Let me know if a visual like this helps explain the concept! I can also work on more complex graphs depicting reaction rates, equilibrium etc. if helpful.*

At this point, the student asked the GenAIbots for a more detailed illustration of these concept. In response, **ChatGPT** effectively leveraged its *Advanced Data Analysis* mode to provide a graph that directly addressed the student's inquiry, elucidating the dynamics of the exemplified chemical reaction over time, demonstrating excellence in the **Instructional Strategies** and **Detail Depth & Content Recommendation** categories. Its graph showcased the decline in reactant concentrations alongside the rise in product concentration over time, making it highly relevant and pedagogically valuable.

Conversely, **BingChat**, despite having the advantage of direct access to the DALL-E 3 image generator (David, 2023), failed to provide a personalized response to the student's request. While the generated images might have been undeniably high quality, the lack of direct relevance to the student's question and the generic nature of the images suggests shortcomings in the **Personalization & Personable Touch** and **Instructional Strategies** categories. Despite this, this integration is a welcomed step forward, particularly when reflecting on the limitations of the standalone DALL-E 2 version in tasks like lettering, as it is evident in the example below that resulted from the student asking Bing Image Creator to illustrate the concept of chemical reactions.

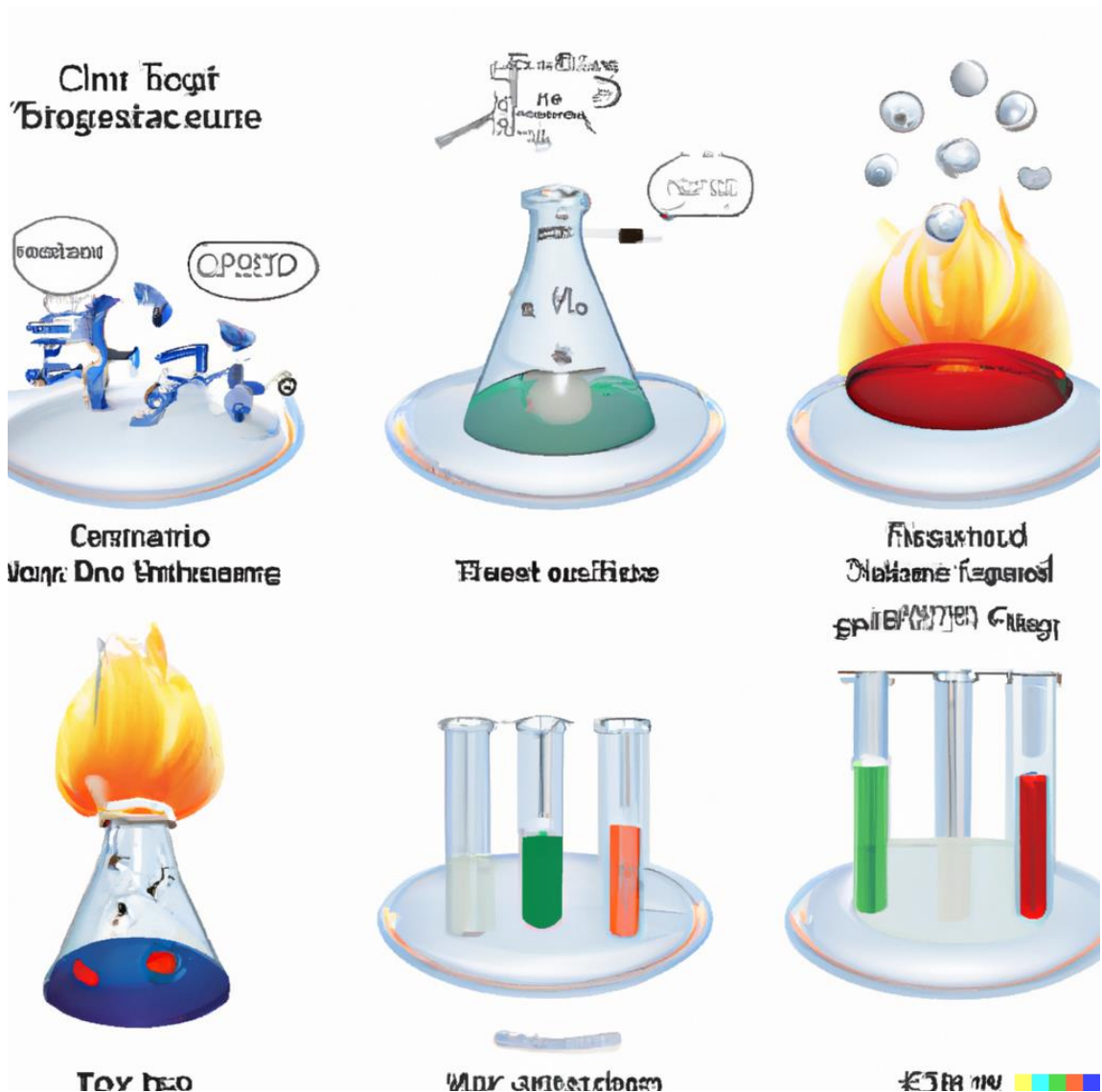

Lacking the capability to generate images, **Bard** showcased its adaptability by sourcing available online images that depicted graphs of water's phase changes. However, this method inherently risks not being precisely tailored to the question at hand. Even though it might have tapped into **Detail Depth & Content Recommendation**, the reliance on pre-existing content might not always be the most effective way to address specific inquiries. **Claude** did attempt to employ **Instructional Strategies** by executing an internal code to generate a graph showcasing the dynamics of the chemical reaction over time, but the visual representation was of inferior quality, possibly confusing rather than clarifying, and weakening the effectiveness of the **Engagement & Interactivity** and **Use of Analogy & Comparative Illustrations** categories.

It is worth noting that DALL-E 3 is also integrated into ChatGPT (Metz & Hsu, 2023) but as a distinct mode separate from the *Advanced Data Analysis* mode utilised in this session. To access its capabilities, the student would need to initiate a new chat session in that mode and reiterate the question. However, the images produced in this mode would be much more closely aligned with the student's query, as seen below.

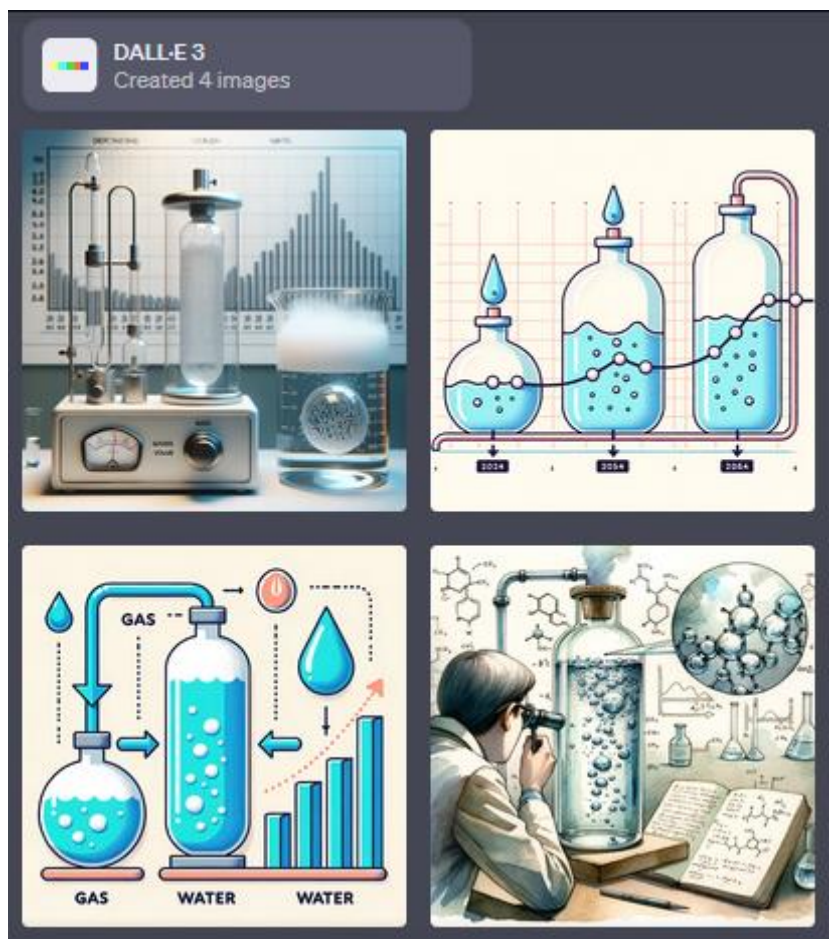

## Question about the difference between atoms, molecules, and moles

**P1:** ah. this thing also confuses me. what is the difference between atom, molecule and mole?
**ChatGPT:** <u>Certainly, it's a common area of confusion for many students</u>. <u>Let's break it down one by one</u>, <u>using simple analogies</u>:
1. **Atom:**
   - *Imagine an atom as a single LEGO brick. It's the tiniest piece of an element that still retains the properties of that element. Just like a single red LEGO brick represents the color red, a single hydrogen atom represents the element hydrogen.*
2. **Molecule:**
   - *When you connect several LEGO bricks together, you form a structure. In chemistry, when you connect several atoms together, you form a molecule. For instance, two hydrogen atoms (H) can connect to form a hydrogen molecule ($H_2$). Similarly, two hydrogen atoms and one oxygen atom can connect to form a water molecule ($H_2O$).*
3. **Mole:**
   - *This one's a bit more abstract. Let's say you have a big box filled with identical LEGO structures (say, LEGO houses). If you count them and find there are exactly $6.022 \times 10^{23}$ LEGO houses in that box, you have a "mole" of LEGO houses.*

- In chemistry, a mole is a specific number of things (atoms, molecules, electrons, etc.). Specifically, one mole of any substance contains exactly 6.022 x $10^{23}$ of those entities. This number is called Avogadro's number.
- The concept of a mole is a standardized way to talk about quantities in chemistry, much like a "dozen" is a standardized way to talk about 12 of something.

So, to recap:
- **Atom:** Smallest unit of an element.
- **Molecule:** Made up of multiple atoms bonded together.
- **Mole:** A specific quantity (6.022 x $10^{23}$) of atoms, molecules, or other entities.

Do these analogies help clarify the differences for you?

**BingChat:** I see that you are confused about the difference between atom, molecule and mole. These are three important terms in chemistry that describe different aspects of matter. Let me try to explain them to you in a simple way.

An atom is the smallest unit of an element that retains its chemical identity. An element is a pure substance that consists of only one type of atom. For example, hydrogen, oxygen, carbon, and gold are elements. Each element has a unique symbol and atomic number that indicates how many protons it has in its nucleus. For example, hydrogen has the symbol H and atomic number 1, meaning it has one proton in its nucleus. Atoms are composed of subatomic particles such as protons, neutrons, and electrons. Protons and neutrons are found in the nucleus, and electrons orbit around the nucleus in shells or orbitals. The number of electrons in an atom determines its chemical behavior and reactivity [12].

A molecule is a group of two or more atoms that are held together by chemical bonds. A chemical bond is a force of attraction between atoms that share or transfer electrons. There are different types of chemical bonds, such as covalent, ionic, metallic, and hydrogen bonds. A molecule can be composed of atoms of the same element or different elements. For example, oxygen gas (O2) is a molecule composed of two oxygen atoms bonded by a covalent bond. Water (H2O) is a molecule composed of two hydrogen atoms and one oxygen atom bonded by covalent bonds. A molecule has a specific shape and structure that determines its physical and chemical properties [32].

A mole is a unit of measurement that is used to count the number of atoms or molecules in a given amount of substance. A mole is defined as the amount of substance that contains as many particles (atoms or molecules) as there are atoms in 12 grams of carbon-12. This number is called Avogadro's number and it is approximately 6.022 x $10^{23}$. That means one mole of any substance contains 6.022 x $10^{23}$ particles (atoms or molecules). For example, one mole of oxygen gas contains 6.022 x $10^{23}$ molecules of O2, and one mole of water contains 6.022 x $10^{23}$ molecules of H2O. The mass of one mole of a substance is called its molar mass and it is equal to its molecular weight in grams. For example, the molecular weight of oxygen gas is 32 g/mol and the molecular weight of water is 18 g/mol [42].

So, to summarize, an atom is the smallest unit of an element, a molecule is a group of atoms bonded together, and a mole is a unit of measurement that counts the number of particles in a substance. I hope this helps you understand the difference between atom, molecule and mole.

**Bard:** An atom is the smallest unit of matter that still has all of the properties of its element. An element is a substance that cannot be broken down into simpler substances by chemical means.

A molecule is a group of two or more atoms held together by chemical bonds. A chemical bond is a force that holds atoms together.

*A mole is a unit of measurement that is used to count atoms and molecules. One mole of any substance contains 6.02214076×10^23 particles of that substance. This number is known as Avogadro's constant.*
*Here is a table that summarizes the key differences between atoms, molecules, and moles:*

| Characteristic | Atom | Molecule | Mole |
|---|---|---|---|
| Size | Smallest unit of matter | Group of two or more atoms | Unit of measurement |
| Composition | Made up of protons, neutrons, and electrons | Made up of two or more atoms | Can be made up of any type of atom |
| Number of particles | One | Two or more | 6.02214076×10^23 |

*Here are some examples of atoms, molecules, and moles:*
- *Atoms: Hydrogen (H), helium (He), oxygen (O), carbon (C)*
- *Molecules: Water (H2O), carbon dioxide (CO2), methane (CH4)*
- *Moles: One mole of hydrogen atoms, one mole of water molecules, one mole of carbon dioxide molecules*

*I hope this helps to explain the difference between atoms, molecules, and moles. Please let me know if you have any other questions.*

**Claude:** *Excellent question! Those three terms can definitely be confusing. Let's go through them one-by-one:*

*Atom - The basic building block of matter. Atoms are made up of protons, neutrons, and electrons. An atom is the smallest unit of an element that maintains the properties of that element. For example, a hydrogen atom has 1 proton; a carbon atom has 6 protons.*

*Molecule - A group of two or more atoms bonded together. A molecule is the smallest unit of a chemical compound that maintains the properties of that compound. For example, a water molecule contains 2 hydrogen atoms and 1 oxygen atom bonded together (H2O).*

*Mole - This is a number representing 6.02 x 10^23 things. It's usually used to represent a certain quantity of atoms or molecules. For example, 1 mole of carbon contains 6.02 x 10^23 carbon atoms.*

*So in summary:*
- *Atoms are the basic units of elements*
- *Molecules are groups of bonded atoms that make up compounds*
- *Moles are a way to represent 6.02 x 10^23 of something (useful for relating atomic/molecular masses to macroscopic masses)*

*Let me know if this helps explain the distinction between these three important chemistry concepts! I'm happy to provide more examples as well.*

**ChatGPT**, by saying "Certainly, it's a common area of confusion for many students", recognises a **Reinforcement & Feedback** element by acknowledging the difficulty many students face. By saying, "Let's break it down one by one," it demonstrates **Instructional strategy**. The phrase "using simple analogies" clearly indicates the use of **Analogy & Comparative**

**Illustrations**, while the question "Do these analogies help clarify?" shows an **Engagement & interactivity** component by asking for feedback. **BingChat** saying "I see that you are confused" provides **Reinforcement & Feedback**, while the phrase "Let me try to explain" indicates an **Instructional Strategy** and "in a simple way" attempts at **Detail Depth & Content Recommendation** by simplifying the information. Finally, "I hope this helps" and the inclusion of the smiley emoticon "😊" are gestures for **Personalization & Personable Touch**. **Bard**, by explaining "An atom is..." and, later, offering "Here is a table," appeals to an **Instructional Strategy,** demonstrated by definition-based learning and visual data organisation, concluding with "I hope this helps" and "Please let me know if you have any other questions" demonstrates **Engagement & Interactivity** and **Reinforcement & Feedback**. **Claude**, by starting with "Excellent question" and acknowledging that the subject "can definitely be confusing," recognises gives **Reinforcement & feedback**. It then invites, "Let's go through them one-by-one," pointing to an **Instructional Strategy** through systematic explanation. It also uses **Analogy & Comparative Illustrations** by providing two examples. It concludes with "Let me know if this helps," an **Engagement & Interactivity** element, and encouraging feedback.

**Deepening on the concept of mole**

*P1: so can you talk about a mole of beer bottles?*
*ChatGPT: Absolutely! The concept of a "mole" isn't exclusive to atoms or molecules; it's simply a specific quantity. In the realm of chemistry, one mole always refers to 6.022 x 10$^{23}$ of some entity, whether that's atoms, molecules, electrons, or even beer bottles.*
*So, when we say "a mole of beer bottles," we're referring to 6.022 x 10$^{23}$ beer bottles. That's a LOT of beer bottles!*
*However, it's worth noting that outside of a chemistry context, using "mole" to refer to quantities of everyday items might be confusing. In everyday language, we have other terms for large quantities, like "dozen" (for 12 of something) or "gross" (for 144 of something). But theoretically, yes, you can have a mole of anything, including beer bottles.*
*Just to give you a sense of how large this number is: If you were to try and count to Avogadro's number at a rate of one number per second, it would take you more than 19 billion years!*
*Would you like to explore more about the implications of having a mole of beer bottles or any other fun examples?*
*BingChat: Yes, I can talk about a mole of beer bottles, but it might not be very practical. A mole of beer bottles is a very large number of beer bottles, much more than you can ever drink or see in your lifetime. Let me explain why.*
*A mole of beer bottles is the amount of beer bottles that contains as many bottles as there are atoms in 12 grams of carbon-12. This number is approximately 6.022 x 10$^{23}$, which is a very big number. To give you some idea of how big this number is, here are some comparisons:*
  - *If you had a mole of beer bottles, you could cover the entire surface of the Earth with about 100 layers of beer bottles.*

- If you had a mole of beer bottles, you could <u>fill up the Pacific Ocean</u> with beer bottles about 30 times.
- If you had a mole of beer bottles, you could stack them up to reach the moon and back about 2000 times.

As you can see, a mole of beer bottles is a huge amount of beer bottles that is impossible to handle or imagine. That's why chemists use the mole as a unit of <u>measurement for atoms and molecules</u>, which are very small and numerous, rather than for everyday objects like beer bottles. A mole of atoms or molecules has a mass that is easy to measure and compare with other substances. <u>For example</u>, one mole of oxygen gas has a mass of 32 grams, and one mole of water has a mass of 18 grams.

So, to summarize, a mole of beer bottles is a hypothetical amount of beer bottles that contains 6.022 x 10$^{23}$ bottles. This number is so large that <u>it is not useful or meaningful for describing real-world situations</u>. A mole is more suitable for measuring the amount of atoms or molecules in a substance. <u>I hope this helps</u> you understand what a mole of beer bottles means. 😊

**Bard:** <u>Yes, I can talk about a mole of beer bottles.</u>

A mole of beer bottles is a unit of measurement that is used to count beer bottles. One mole of beer bottles contains 6.02214076×10^23 beer bottles. This number is known as Avogadro's constant.

<u>To put this into perspective</u>, a mole of beer bottles is a lot of beer bottles! It would take over <u>100 billion years</u> to count them all, at a rate of one bottle per second.

Of course, no one has ever seen a mole of beer bottles in real life. It is a <u>theoretical concept</u> that is <u>used by chemists</u> to count atoms and molecules.

But that doesn't mean that we can't have fun thinking about it. <u>What would a mole of beer bottles look like</u>? How much would it weigh? How much beer would it contain?

If we assume that each beer bottle is 12 ounces, then a mole of beer bottles would weigh over 100 million pounds. And it would contain over <u>700 million gallons of beer</u>!

That's a lot of beer!

So, <u>next time you're enjoying a cold beer with your friends</u>, take a moment to think about the mole of beer bottles that it represents. It's a truly <u>mind-boggling</u> concept.

**Claude:** Sure, I'm happy <u>to explain the concept of a mole using beer bottles as an example</u>.

A mole represents a <u>specific number</u> - 6.02 x 10^23 - <u>of anything</u>, whether atoms, molecules, <u>or objects like beer bottles</u>.

<u>Let's imagine a mole of beer bottles</u>. That would be 6.02 x 10^23 individual bottles. That's <u>a huge number of bottles</u>!

<u>To put it in perspective</u>:
- If you had 1 mole of beer bottles and stacked them on top of each other, the stack would reach about <u>1.5 billion light years tall</u> - that's halfway across the observable universe!
- If you lined up 1 mole of beer bottles side by side, the line would stretch for 9 trillion kilometers - over <u>200 times the distance between Earth and Pluto</u>.
- If you purchased 1 mole of beer bottles at $10 per bottle, it would <u>cost 60.2 sextillion dollars</u> - way more than the GDP of Earth!

So a mole represents an enormous number of things. It's <u>very useful in chemistry</u> for relating atomic/molecular masses to measurable macroscopic masses. For example, if you know the mass of 1 mole of carbon atoms, you can use that to calculate how many carbon atoms are in a certain macroscopic sample.

<u>Hope this silly example</u> with beer bottles helps give you an idea of just how huge a mole really is! <u>Let me know if you need any clarification</u> or want <u>to go through more practice problems</u> applying the mole concept.

**ChatGPT** starts with "Absolutely!" offering **Personalization & Personable Touch**. It then discusses the "mole" concept using **Instructional Strategies** by noting that while it is a "specific quantity" in the "realm of chemistry," it is not exclusively tied to "atoms" or "molecules" and enters a **Meta-discussion & Perspective Framing** by conceding that "outside of a chemistry context," the general use of the word 'mole' "might be confusing". Proceeding further, the explanation by making analogies with everyday terms like "dozen" and "gross", also with the **Use of Analogy & Comparative Illustrations**. The perspective on counting to Avogadro's number "to give a sense" of how large this number is and the final invitation to discuss "implications" or "any other fun examples" leans into **Engagement & Interactivity**. **BingChat** also starts by offering **Personalization & Personable Touch,** conceding that "Yes", one "can talk about a mole of beer bottles" but right after turns to **Instructional Strategies**, by referring that "it might not be very practical" and later that "it is not useful or meaningful for describing real-world situations". It then provides a detailed explanation of the mole concept, highlighting the enormousness of "6.022 x 10^23" through real-world scenarios like covering the "Earth" or filling the "Pacific Ocean", which emphasises the **Use of Analogy & Comparative Illustrations**. Categorisation of the mole as primarily for counting "atoms/molecules" again showcases **Instructional Strategies**, while providing an example of the application of the concept to oxygen gas aligns with the **Use of Analogy & Comparative Illustrations**. The closing "I hope this helps" followed by an emoji implies a touch of **Personalization & Personable Touch**. **Bard** likewise starts with "Yes, I can talk about a mole of beer bottles," offering **Personalization & Personable Touch**. It then engages the reader with the enormity of "6.02214076×10^23 beer bottles", by undergoing a detailed calculation of "What would a mole of beer bottles? look like", offering perspectives on time ("100 billion years") and volume ("700 million gallons of visualisation and "To put this into perspective", aligning with **Use of Analogy & Comparative Illustrations**. Highlighting it as a "theoretical concept" "used by chemists emphasises **Instructional Strategies** and the jovial remark about enjoying beer with friendly tone emphasising the mind-bogglingness of the indicates **Personalization & Personable Touch**. **Claude** starts with **Instructional Strategies**, accepting beer bottles as an example to breakdown of the mole concept and its application as a "specific number," "very useful in chemistry," that may be applied to count "anything". It then invites one to "imagine a mole of beer bottles", which would be "a ,huge number" and puts "it in perspective", drawing vivid pictures such as a stack reaching "1.5 billion light years", a line stretching "200 times the distance between Earth and Pluto," and the cost of "60.2 sextillion dollars," underscoring the

**Use of Analogy & Comparative Illustrations**. The closing "Hope this silly example" and followed by invitation for more "practice problems" or "clarifications" hints at **Reinforcement & Feedback**. Nevertheless, while the use of the word "silly" by Claude might intend to suggest that the example provided is not to be taken too seriously or is trivial in nature, it also could be perceived as a bit demeaning or dismissive, negatively impacting the **Personalization & Personable Touch** category.

**Question about changing of state**

*P1:* You mentioned water and water vapor. this is what my professor calls a change of state, right?
*ChatGPT:* Yes, exactly! A change of state, also known as a phase transition, refers to the transformation of a substance from one state of matter to another. The most common states of matter are solid, liquid, and gas. The transitions between these states are typically due to changes in temperature and/or pressure.
Let's go over the primary changes of state:
1. **Melting**: The transition from solid to liquid. For instance, when ice (solid water) warms up, it melts to form liquid water.
2. **Freezing:** The transition from liquid to solid. It's the opposite of melting. When liquid water cools down, it freezes to form ice.
3. **Boiling** (**or Vaporization**): The transition from liquid to gas. When water is heated to its boiling point, it turns into water vapor (steam).
4. **Condensation**: The transition from gas to liquid. It's the opposite of boiling. When steam cools down, it condenses to form liquid water.
5. **Sublimation**: The transition directly from solid to gas without passing through the liquid state. Dry ice (solid carbon dioxide) does this—it turns directly into carbon dioxide gas when it warms up.
6. **Deposition**: The transition directly from gas to solid without passing through the liquid state. It's the opposite of sublimation.
Each substance has specific temperatures at which these changes of state occur, called its melting point, freezing point, boiling point, etc. These points can also vary based on external pressure.
Does this explanation help clarify the concept of changes of state for you? Would you like to dive deeper into any specific transition or related topic?
*BingChat:* Yes, you are right. Water and water vapour are examples of a change of state. A change of state is a physical change in which a substance changes from one state of matter to another. The three common states of matter are solid, liquid, and gas. Water can exist in all three states, depending on the temperature and pressure. For example, water is a liquid at room temperature and normal atmospheric pressure, but it can change to a solid (ice) when it is cooled below 0°C, or to a gas (water vapour) when it is heated above 100°C[123].
There are different names for the processes that cause a change of state. For example, when a liquid changes to a gas, it is called vaporization. Vaporization can happen in two ways: evaporation or boiling. Evaporation is when a liquid changes to a gas at the surface, without reaching its boiling point. Boiling is when a liquid changes to a gas throughout the liquid, when it reaches its boiling point. When a gas changes to a liquid, it is called condensation. Condensation is when a gas loses heat and turns into a liquid droplet[123].

*A change of state does not affect the chemical composition of a substance. That means the substance does not change into a new substance with different properties. For example, water and water vapour are both made of the same molecules (H2O), but they have different physical properties such as shape, volume, and density. A change of state can be reversed by changing the temperature or pressure again. For example, water vapour can change back to liquid water by cooling it down[123].*

*I hope this helps you understand what a change of state is and how it happens. If you want to learn more about the changes of state of water and other substances, you can watch this video or read this article. 😊*

**Bard:** *Yes, that is correct. Water and water vapor are two different states of water. Water is a liquid at room temperature, but it can be converted to a gas (water vapor) by heating it. Conversely, water vapor can be converted to a liquid (water) by cooling it.*

*This process of changing the state of a substance is called a change of state. There are three main states of matter: solid, liquid, and gas. Each state has different properties.*

*For example, solids have a definite shape and volume. Liquids have a definite volume, but they take the shape of their container. Gases have neither a definite shape nor a definite volume.*

*When a substance changes state, it absorbs or releases energy. For example, when water vapor condenses to form liquid water, it releases energy. This energy is what causes the air around us to feel warm when it rains.*

*Changes of state are important in many different natural processes, such as the water cycle and weather formation. They are also used in many different industrial processes, such as food processing and medicine.*

*I hope this helps to explain what a change of state is. Please let me know if you have any other questions.*

**Claude**: *Yes, that's correct! When water transitions between liquid water and water vapor, it is undergoing a change of state.*

*The different states of matter are:*
- *Solid - Fixed shape and volume, tightly packed molecules*
- *Liquid - Fixed volume but variable shape, flowing molecules*
- *Gas - Variable volume and shape, molecules move freely*

*Water can exist in all three states:*
- *Ice (solid)*
- *Liquid water*
- *Water vapor (gas)*

*Some key changes of state for water:*
- *Melting - Solid to liquid, ice melts to water*
- *Freezing - Liquid to solid, water freezes to ice*
- *Evaporation - Liquid to gas, water evaporates to vapor*
- *Condensation - Gas to liquid, vapor condenses into water*

*The processes of evaporation and condensation describe water transitioning between the liquid and gas states. This is what produces water vapor.*

*Other examples of changes of state include:*
- *Boiling - Liquid to gas, water boils to steam*
- *Sublimation - Solid to gas, dry ice (solid CO2) sublimates to CO2 gas*

*Let me know if you need any clarification on changes of state or the different phases of matter! Being comfortable with these concepts is important for understanding many chemical processes.*

**ChatGPT** acknowledges the student's understanding with an "Yes, exactly!," offering **Personalization & Personable Touch**. It then directly tackles the student's question about water and water vapor, providing a thorough and structured breakdown of state changes, leaning to **Instructional Strategies** and saying, "Let's go over". It finishes the explanation recurring to **Meta-discussion & Perspective Framing** referring that "Each substance has specific temperatures." It concludes with "Does this explanation help clarify" "for you?" offering **Reinforcement & Feedback** and **Personalization & Personable Touch**, respectively, and adeptly inviting further exploration with "Would you like to dive deeper", what points to **Detail Depth & Content Recommendation**. **BingChat** similarly recognizes the student's comprehension by stating "Yes, you are right," providing **Personalization & Personable Touch**. It not only offers a clear and presentation including examples, what hit the **Detail Depth & Content Recommendation** and **Use of Analogy & Comparative Illustrations** categories, but also integrates external content recommendations, pointing to **Depth & Content Recommendation**. It also adds that "A change of state can be reversed," offering **Meta-discussion & Perspective Framing**. It finalizes with "I hope this helps", and a smiley, signs of **Engagement & Interactivity** and **Personalization & Personable Touch** respectively and "If you want to learn more" "you can watch this video or read this article" pointing to **Detail Depth & Content Recommendation**. **Bard** likewise recognizes the student's grasp by stating, "Yes, that is correct," exhibiting a **Personalization & Personable Touch**. It then provides a concise yet illuminating **Meta-discussion & Perspective Framing** on the implications of state changes being "important in many different natural processes," "water cycle," and "weather formation" and "in many different industrial processes," "food processing" and "medicine", connecting the topic to broader contexts. The explanation concludes with "I hope this helps to explain," showcasing **Engagement & Interactivity**, followed by "Please let me know if you have any other questions," which emphasizes **Personalization & Personable Touch**. Like the earlier GenAIbots, **Claude** begins by recognizing the student's comprehension with "Yes, that's correct!," adding **Personalization & Personable Touch**. It then presents the states of matter and their transitions with clear analogies, leaning to **Instructional Strategies**, and extending the explanation with "other examples of changes of state", such as "Boiling" and "Sublimation," what hit the **Detail Depth & Content Recommendation**, although its depth might be a tad more concise that the previous ones. It then concludes with "Let me know if you need any clarification," which emphasizes **Personalization & Personable Touch** and includes the remark, "Being comfortable with these concepts is important for understanding many chemical processes," aligning with the **Meta-discussion & Perspective Framing** category.

**Question about an animated representation of water molecules undergoing phase changes**

**P1:** hmmm. you made a nice graph before. Could you build something more dynamic than a static diagram, an animated representation of water molecules undergoing phase changes?

**ChatGPT:** Got it! Creating an animated representation of water molecules undergoing phase changes requires defining the behavior of the molecules for each phase and then transitioning between these phases.

I'll follow these steps:
1. Represent the molecules as small circles.
2. Define their behavior in the solid phase.
3. Transition to the melting phase.
4. Define behavior in the liquid phase.
5. Transition to the boiling phase.
6. Define behavior in the gas phase.
7. Compile these animations into a continuous GIF.

[Some internal code working]

Here's the animated representation of water molecules undergoing phase changes:
1. **Solid (Ice)**: Molecules are closely packed and vibrating in place.
2. **Melting**: Molecules start to move more freely but remain close.
3. **Liquid (Water)**: Molecules move more fluidly and slide past each other.
4. **Boiling**: Molecules gain energy and move more rapidly.
5. **Gas (Steam)**: Molecules spread out and move freely and rapidly.

You can download the GIF here. [link provided]

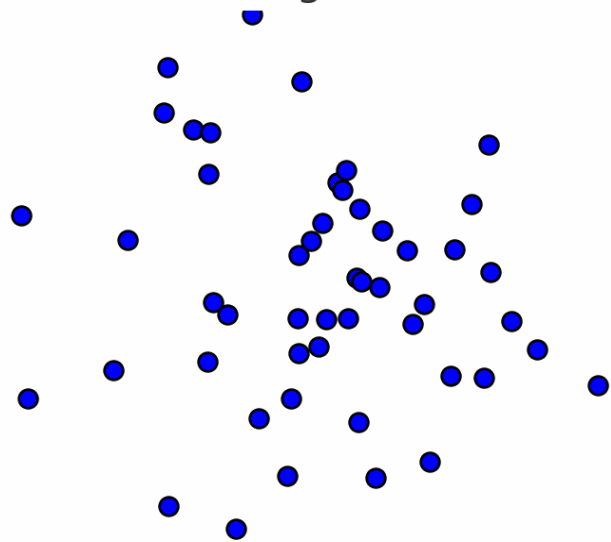

**BingChat:** Sure, I can try to create an animated representation of water molecules undergoing phase changes. I found some web simulations that show how different types of molecules form a solid, liquid, or gas, and how they change their state when heat is added or removed [123]. You can also change the temperature or volume of a container and see a pressure-temperature diagram respond in real time [1]. These simulations might help you understand the molecular behavior of water in different phases.

**Bard:** Sure. Here is a simple animated representation of water molecules undergoing phase changes:

[Animated GIF of water molecules undergoing phase changes]

The animation shows two groups of water molecules. The group on the left is in the liquid phase, and the group on the right is in the gas phase.

The animation starts with the water molecules in the liquid phase. The water molecules are close together and interacting with each other.

As the animation progresses, some of the water molecules in the liquid phase gain enough energy to overcome the forces that are holding them together. These water molecules escape from the liquid phase and form the gas phase.

The water molecules in the gas phase are far apart and not interacting with each other.

The animation also shows how water molecules can condense from the gas phase to the liquid phase. When a water molecule in the gas phase loses energy, it can attract to other water molecules and form a liquid droplet.

The animation can be used to illustrate the different phase changes of water, such as melting, boiling, evaporation, and condensation.

I hope this is helpful!

**Claude:** Sure, animated visualizations can be really helpful for understanding dynamic processes like phase changes! Here is a simple animated illustration showing water molecules in different states:

```
Solid (Ice):
O O
O O
O O
O O

The molecules are stationary in fixed positions (vibrating a little).

Liquid:

O   O O
    O O
 O O

The molecules can slide past each other and move around within the li

Gas (Water Vapor):

    O
   O
  O O
 O   O
  O
 O
```

*In this animation:*
- *The solid ice molecules are stationary and packed together*
- *The liquid molecules flow and move closer together*
- *The gas molecules bounce around freely at large distances*

*To transition between states:*
- *Heating causes more vigorous movement - melting ice or boiling water*
- *Cooling causes slower movement - freezing water or condensing vapor*

*Let me know if an animated illustration like this helps visualize the molecular differences between solids, liquids and gases! I can build more complex animations or use online tools to create interactive phase change simulations. Visuals can really aid understanding of physical chemistry concepts.*

**ChatGPT** was able to recur again to its *Advanced Data Analysis* resources to generate an animated GIF as an **Instructional Strategies**, which, albeit a bit simple and crude, sufficiently illustrated the "water molecules undergoing phase changes" to provide **Engagement & Interactivity**. Furthermore, crafting an original GIF could be seen as **Personalization & Personable Touch**. After promising "to create an animated representation", **BingChat** opted to search the internet, offering links to "some web simulations" that "might help" understand the molecular behavior. This approach aligns with **Instructional Strategies** and **Detail Depth & Content Recommendation**. However, by leaning on pre-existing resources rather than crafting custom content, it could lessen the immediate **Engagement & Interactivity** and potentially diminish the **Personalization & Personable Touch**. **Bard** made an assurance to deliver "a simple animated representation" as a commitment to **Instructional Strategies**, **Personalization & Personable Touch** and **Engagement & Interactivity**, but this ultimately falls short as neither a visible animation nor a download link for such a GIF was provided. Meanwhile, while **Claude** acknowledges that "Visuals can really aid understanding" and "animated visualizations can be really helpful" as **Instructional Strategies**, its attempt was limited to the of a static image, devoid of the requested animation or dynamic representation and lessening **Engagement & Interactivity** and **Detail Depth & Content Recommendation**.

**Question about plasma, a state of matter**

*P1: my teacher said that plasma is the fourth state of matter. but isn't plasma that thing they inject into a vein in the hospital?*
***ChatGPT:*** *<u>Your teacher is correct</u>, and <u>I understand the confusion</u>. The term "plasma" is used in both physics and medicine, but they refer to entirely different concepts. <u>Let's break them down</u>:*
1. ***Plasma (State of Matter):***
- *In physics, plasma is often referred to as the "fourth state of matter" after solids, liquids, and gases.*
- *Plasma is like a gas but with some key differences. In plasma, a significant number of atoms are ionized, meaning they have lost or gained electrons. This gives plasma unique properties, like the ability to conduct electricity and respond strongly to electromagnetic fields.*
- *<u>Examples of plasmas</u> include the <u>sun, lightning, and neon signs</u>. The sun is essentially a giant ball of plasma, with hydrogen atoms undergoing fusion to form helium, releasing a tremendous amount of energy in the process.*
2. ***Plasma (Biological):***
- *In a biological or medical context, plasma refers to the clear, yellowish fluid component of blood. It's what's left after you remove the red and white blood cells and platelets from blood.*
- *Plasma makes up about 55% of total blood volume and contains water, salts, enzymes, hormones, and waste products. It serves as a transport medium for delivering nutrients to the cells and carrying waste products away.*
- *When <u>people donate "plasma" at a donation center</u>, they're donating this component of their blood.*

*So, while the terms are the same, the contexts are quite different. It's essential to know which "plasma" is being referred to based on the topic of discussion.*
*Does this distinction between the two concepts of plasma <u>clear things up for you?</u>*
***BingChat:*** *<u>That's a very good question</u>. Plasma is indeed the fourth state of matter, but it is also a term used in medicine to describe the liquid part of blood. <u>Let me explain the difference</u> between these two types of plasma.*

*Plasma as a state of matter is a gas that has been heated or electrified to the point where some or all of its atoms lose their electrons. This creates a mixture of positively charged ions and negatively charged electrons that can conduct electricity and respond to magnetic fields. Plasma is very common in the universe, as it makes up most of the stars, including the sun. Plasma is also found in lightning, neon lights, and plasma displays[123].*

*Plasma as a part of blood is a yellowish fluid that carries water, salts, enzymes, proteins, hormones, and other substances throughout the body. It also helps with immunity, blood clotting, blood pressure, and pH balance. Plasma can be donated to help people who have lost blood due to injury, surgery, or disease. Plasma can also be used to make medicines for people who have certain conditions that affect their blood[456].*

*So, as you can see, plasma can mean different things depending on the context. Both types of plasma are important for life and science, but they are not the same thing. I hope this helps you understand the difference between plasma as a state of matter and plasma as a part of blood. 😊*

**Bard:** *Yes, plasma is the fourth state of matter. It is a highly energized state of matter in which some or all of the atoms have been stripped of one or more of their electrons. This results in a gas of positively charged ions and negatively charged electrons.*

*Plasma is the most abundant form of matter in the universe, accounting for over 99% of the visible universe. It is found in stars, nebulae, and other celestial objects. Plasma is also found on Earth, in lightning bolts, auroras, and other natural phenomena.*

*Plasma is also used in a variety of industrial applications, such as welding, plasma cutting, and semiconductor manufacturing.*

*The plasma that is injected into veins in the hospital is a type of blood plasma. Blood plasma is the liquid portion of blood that remains after the red blood cells, white blood cells, and platelets have been removed. It is a yellowish liquid that contains proteins, antibodies, and other nutrients.*

*Blood plasma is used to treat a variety of medical conditions, including dehydration, malnutrition, and shock. It is also used to make plasma products, such as albumin and immunoglobulin.*

*So, to summarize, plasma is both the fourth state of matter and a type of blood plasma. The two are not related, but they both have important roles.*

*I hope this helps to clarify things. Please let me know if you have any other questions.*

**Claude:** *Great observation! Plasma does indeed have two different meanings in science:*

1. *In chemistry and physics, plasma is considered the fourth state of matter, distinct from solids, liquids and gases. Plasma consists of highly ionized gas containing a mix of free electrons, positive ions, and neutral atoms. It can be created by heating a gas to very high temperatures. Natural examples include lightning and the Sun.*
2. *In medicine, plasma refers to the liquid portion of blood that contains proteins, nutrients, and ions like sodium and chloride. This plasma carries blood cells and helps regulate bodily processes. It makes up about 55% of total blood volume.*

*So in summary:*
- *Scientific plasma is a distinct state of ionized gas, created at extremely high temps.*
- *Medical plasma is the liquid part of blood that suspends blood cells and transports useful substances.*

*While both are important in their respective fields, they have very different definitions and properties. The medical usage of the term is probably more familiar to most people. But the concept of plasma as a state of matter is critical in physics and chemistry when studying things like nuclear fusion reactions.*

*<u>Let me know</u> if this explanation helps differentiate the two meanings! <u>It's an important distinction</u> to understand.*

    **ChatGPT** starts reassuringly with "Your teacher is correct" and "I understand the confusion," providing **Personalization & Personable Touch**. It then proceeds to **Instructional Strategies**, addressing the need to differentiate the two meanings systematically with "Let's break them down" and expounding extensively a clear distinction between plasma as the fourth state of matter and blood plasma in the medical context with **Detail Depth & Content Recommendation**, as well as with the **Use of Analogy & Comparative Illustrations**, recurring to real-world examples to explain the concept, like the "sun, lightning, and neon signs" as well as of people donating "'plasma' at a donation center". It ends the explanation with **Engagement & Interactivity** by asking "Does this distinction ... clear things up for you?", inviting a response. **BingChat** also starts by giving positive feedback and personal recognition of the user's query with "That's a very good question," providing **Personalization & Personable Touch**. It then proceeds to **Instructional Strategies** with "Let me explain the difference," setting the tone for a structured explanation with **Meta-discussion & Perspective Framing** strategy that contextualizes the significance of both plasmas in the universe and medical science, the **Use of Analogy & Comparative Illustrations**, referencing the sun, "lightning, neon lights, and plasma displays", and links to external resources, offering **Detail Depth & Content Recommendation**. It ends with **Reinforcement & Feedback**, emphasizing the significance of the dual meaning of the word, by affirming that "Both types of plasma are important for life and science" and "I hope this helps you understand the difference" followed by an emoticon for a friendly touch of **Engagement & Interactivity**. **Bard** begins its **Instructional Strategies** with the straightforward "Yes" affirmation followed by a detailed distinction of both concepts with **Detail Depth & Content Recommendation**, progressing from the **Use of Analogy & Comparative Illustrations** with examples of "lightning bolts, auroras, and other natural phenomena" to a **Meta-discussion & Perspective Framing** discussing the broader applications "in a variety of industrial applications, such as welding, plasma cutting, and semiconductor manufacturing", as well "to treat a variety of medical conditions, including dehydration, malnutrition, and shock" to make products, "such as albumin and immunoglobulin". It ends with **Engagement & Interactivity** in "Please let me know if you have any other questions" and **Personalization & Personable Touch** with "I hope this helps to clarify things." **Bard** initiates its response with a straightforward affirmation, "Yes", demonstrating its **Instructional Strategies**. It then elaborates in depth about the concepts, highlighting **Detail Depth & Content Recommendation**. Utilizing the **Use of Analogy & Comparative Illustrations**, it gives examples like "lightning bolts, auroras, and other natural phenomena". Venturing into **Meta-discussion & Perspective Framing**, it underlines the broader applications in sectors such as "welding, plasma cutting, and semiconductor manufacturing", and the medical utility "to treat a variety of medical conditions, including dehydration, malnutrition, and shock" and the creation of products like "albumin and immunoglobulin". Wrapping up, it imparts a **Personalization & Personable Touch** with "I hope this helps to clarify things" and fosters **Engagement & Interactivity** with "Please let me know if you have any other questions." **Claude** also appreciates the user's observation with "Great observation!" demonstrating **Personalization & Personable Touch**. It then turns to **Instructional Strategies** to make a clear distinction between the concepts, utilizing the **Use of Analogy & Comparative Illustrations** by giving the examples of "lightning" and the "Sun". Expanding into **Meta-discussion & Perspective Framing**, it underlines the concept "is critical in physics and chemistry" with applications to "nuclear fusion reactions". It ends with a **Personalization & Personable Touch** with "Let me know if this explanation helps" and provides **Meta-discussion & Perspective Framing**, by affirming that "It's an important distinction to understand."

**Question about chemical bondings**

*P1: I also get confused with covalent bonding, ionic bonds and hydrogen bonds*
*ChatGPT: <u>No worries!</u> The different types of chemical bonds <u>can be a bit confusing initially</u>, but let's use some <u>analogies and examples</u> to clarify the distinctions.*
1. ***Covalent Bonding:***
    - *<u>Analogy</u>: Think of covalent bonds as <u>two people holding hands</u>, where each person represents an atom. Both people are sharing their hands (or electrons) with each other.*
    - *In covalent bonding, two atoms share one or more pairs of electrons. This sharing helps both atoms achieve a more stable electron configuration (usually a full outer shell).*
    - *<u>Example</u>: The bond between hydrogen and oxygen in a water molecule ($H_2O$) is covalent.*
2. ***Ionic Bonding:***
    - *<u>Analogy</u>: Imagine a scenario where <u>one person gives a gift (an electron) to another person</u>. The person who gives the gift now feels a sense of loss, while the recipient feels a sense of gain.*
    - *In ionic bonding, one atom donates an electron to another atom. This creates two ions: a positively charged cation (the donor) and a negatively charged anion (the recipient). Because opposite charges attract, these two ions are held together by electrostatic forces.*
    - *<u>Example</u>: In table salt (NaCl), sodium ($Na^+$) donates an electron to chlorine ($Cl^-$), forming an ionic bond.*
3. ***Hydrogen Bonds:***

- *Analogy: Think of hydrogen bonds as weaker handshakes between neighbors. They're not as strong or intimate as holding hands, but they still create a connection.*
- *Hydrogen bonds are intermolecular forces (meaning they occur between molecules) and are weaker than covalent or ionic bonds. They form when a hydrogen atom bonded to a highly electronegative atom (like oxygen or nitrogen) is attracted to another electronegative atom.*
- *Example: In water, the hydrogen atoms of one water molecule are attracted to the oxygen atom of a neighboring water molecule, forming a hydrogen bond. This is why water has unique properties like a high boiling point.*

Here's a brief recap:
- **Covalent Bonds:** Atoms share electrons.
- **Ionic Bonds:** Atoms transfer electrons, resulting in attraction between positive and negative ions.
- **Hydrogen Bonds:** Weak attractions between molecules involving a hydrogen atom.

Does this help clarify the differences for you? Would you like to explore any specific type of bond in more detail?

**BingChat:** *I understand that covalent bonding, ionic bonds and hydrogen bonds can be confusing, as they are different types of interactions between atoms or molecules. Let me try to explain them to you in a simple way.*

*Covalent bonding is a type of intramolecular force, which means it occurs within a molecule. Covalent bonding is when two atoms share one or more pairs of electrons to achieve a stable configuration. For example, in a water molecule, each hydrogen atom shares one electron with the oxygen atom, forming a covalent bond. Covalent bonds can be single, double, or triple, depending on how many pairs of electrons are shared. Covalent bonds are usually strong and require a lot of energy to break[1][2].*

*Ionic bonds are another type of intramolecular force, which means they also occur within a molecule. Ionic bonds are when one atom transfers one or more electrons to another atom, creating oppositely charged ions that attract each other. For example, in a sodium chloride crystal, each sodium atom loses one electron to become a positive ion (cation), and each chlorine atom gains one electron to become a negative ion (anion). The cations and anions are held together by ionic bonds. Ionic bonds are also strong and require a lot of energy to break[1][2].*

*Hydrogen bonds are a type of intermolecular force, which means they occur between molecules. Hydrogen bonds are when a hydrogen atom that is covalently bonded to an electronegative atom (such as oxygen, nitrogen, or fluorine) forms a weak attraction with another electronegative atom on a different molecule. For example, in liquid water, each hydrogen atom on one water molecule can form a hydrogen bond with the oxygen atom on another water molecule. Hydrogen bonds are much weaker than covalent or ionic bonds and can be easily broken by thermal energy[1][2].*

*I hope this helps you understand the difference between covalent bonding, ionic bonds and hydrogen bonds. If you want to learn more about these types of bonds and see some examples, you can visit this website[1] or watch this video[3]. 😊*

**Bard:** *Covalent bonds, ionic bonds, and hydrogen bonds are all types of chemical bonds. A chemical bond is a force that holds atoms together.*

*Covalent bonds are formed when two atoms share one or more pairs of electrons. Covalent bonds are typically formed between nonmetals. Examples of covalent bonds include the bonds in water (H2O), methane (CH4), and carbon dioxide (CO2).*

*Ionic bonds are formed when one atom transfers one or more electrons to another atom. This results in two oppositely charged ions, which are attracted to each other by electrostatic forces.*

*Ionic bonds are typically formed between metals and nonmetals. Examples of ionic bonds include the bonds in sodium chloride (NaCl) and calcium chloride (CaCl2).*

*Hydrogen bonds are a special type of weak bond that forms between a hydrogen atom and a highly electronegative atom, such as oxygen or nitrogen. Hydrogen bonds are responsible for many of the unique properties of water, such as its high boiling point and surface tension.*

*Here is a table that summarizes the key differences between covalent bonds, ionic bonds, and hydrogen bonds:*

| Characteristic | Covalent bond | Ionic bond | Hydrogen bond |
|---|---|---|---|
| Type of bond | Sharing of electrons | Transfer of electrons | Weak bond between hydrogen and electronegative atom |
| Formed between | Nonmetals | Metals and nonmetals | Hydrogen and electronegative atom |
| Examples | Water (H2O), methane (CH4), carbon dioxide (CO2) | Sodium chloride (NaCl), calcium chloride (CaCl2) | Water (H2O), ammonia (NH3), DNA |

**Claude:** *No problem, those bonding terms can definitely be confusing! Let's go through them one-by-one:*
- *Covalent bonds - Shared pairs of electrons between atoms. The electrons are attracted to the nuclei of both atoms, holding them together. Covalent bonds form molecules by connecting atoms. Example: the H2 molecule, with a shared pair of electrons between the H atoms.*
- *Ionic bonds - Electrostatic attraction between positively and negatively charged ions. The ions form from electron transfer between atoms. Ionic bonds hold ionic compounds together in a lattice. Example: table salt (NaCl), formed from Na+ and Cl- ions attracting.*
- *Hydrogen bonds - A special type of dipole-dipole attraction between molecules. It occurs when hydrogen is bonded to highly electronegative atoms (F, O, N) which draws the electron density away. This creates a slightly positive H that can attract the negative end of another molecule. Example: hydrogen bonding between water molecules, which allows ice to maintain structure.*

*The key differences:*
- *Covalent - shared electrons*
- *Ionic - electron transfer*
- *Hydrogen - electrostatic between polar molecules*

*Let me know if you need any clarification or want to go through examples contrasting the different types! Being able to distinguish them will help a lot when learning about chemical bonding.*

  **ChatGPT** starts using **Reinforcement & Feedback**, with a "No worries!," and by assuring that these concepts "can be a bit confusing initially". It then employs **Instructional Strategies** by breaking down the content into bullet points, which makes it more digestible and making **Use of**

**Analogy & Comparative Illustrations** with vivid and relatable examples and analogies like "holding hands," giving "a gift … to another person," and "weaker handshakes between neighbors" to elucidate the nuances of bonding and the importance of these bonds in understanding the properties of compounds in chemistry. The ending questions "Does this help clarify" and "Would you like to explore" in more detail promote a **Personalization & Personable Touch** and **Engagement & Interactivity**, respectively. **BingChat**, on the other hand, employs **Instructional Strategies,** by providing a concise explanation, **Analogy & Comparative Illustrations** with relatable examples, and **Detail Depth & Content Recommendation** providing links to websites. It also uses a **Personalization & Personable Touch** by saying "I hope this helps" and adding an emoticon at the end and further **Detail Depth & Content Recommendation** with "If you want to learn more" and links to websites and videos. **Bard** offers structured, fact-based information and a few examples, in a more conventional classroom style, with tabulation, representing ideal **Instructional Strategies** for learners preferring straightforward factual presentations. However, this approach may lack the **Engagement & Interactivity** other platforms provide. **Claude**, after an encouraging "No problem" and acknowledging that the subject "can definitely be confusing," offering **Reinforcement & Feedback**, takes a more direct approach, also offering structured, fact-based information and a few examples, in a more conventional classroom style, also incurring in the risk of lacking **Engagement & Interactivity**. with phrases that beckon the learner to seek further clarification or contrastive examples. The ending question "Let me know if you need any clarification" promote a **Personalization & Personable Touch** and the invitation to ask for more "examples contrasting the different types" of bonding indicates **Detail Depth & Content Recommendation** and **Engagement & Interactivity**. The concluding remark, "Being able to distinguish them will help a lot when learning about chemical bonding," underscores the wider importance of grasping the distinction under discussion and aligns with the **Meta-discussion & Perspective Framing**".

**Question about illustration of chemical bonds**

*P1:* *can you create an illustration explaining these types of chemical bonds?*
*ChatGPT:* <u>Certainly!</u> <u>Let's break down</u> the explanation and illustration into three parts:
1. **Covalent Bonding**
2. **Ionic Bonding**
3. **Hydrogen Bonding**

*1. Covalent Bonding:*

**Description**: Covalent bonds are formed when two atoms share electrons. The shared electrons are attracted to the nuclei of both atoms, holding the atoms together. This type of bond is common between non-metals.
**Illustration**: Imagine two circles (representing atoms) with electrons (smaller circles or dots) around them. When they come close, they share some of these electrons.
## 2. Ionic Bonding:
**Description**: Ionic bonds are formed when one atom donates one or more electrons to another atom. This creates a positive ion (cation) and a negative ion (anion). The opposite charges attract each other, holding the atoms together. This type of bond is common between metals and non-metals.
**Illustration**: Imagine two circles (representing atoms). One atom has an extra electron which it donates to the other. This creates a "+" sign on the donating atom (cation) and a "-" sign on the receiving atom (anion). They are attracted due to opposite charges.
## 3. Hydrogen Bonding:
**Description**: Hydrogen bonds are a type of weak bond that occurs when a hydrogen atom bonded to a strongly electronegative atom (like oxygen or nitrogen) is attracted to another electronegative atom. They are especially important in water molecules and in the structure of DNA.
**Illustration**: Imagine a water molecule (H2O) with a hydrogen atom (H) and an oxygen atom (O). The hydrogen atom in one water molecule is attracted to the oxygen atom in a neighboring water molecule. This attraction is represented by a dotted line, indicating the hydrogen bond.
Now, _let's create a visual representation_ for each of these bond types.

Covalent Bonding  Ionic Bonding  Hydrogen Bonding

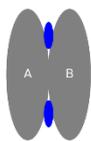 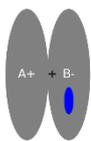 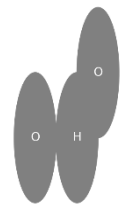

_Here's the visual representation of the three types of bonds:_
1. **Covalent Bonding**: Atoms A and B are sharing electrons, represented by the blue dots between them. The red line indicates the covalent bond formed by this sharing.
2. **Ionic Bonding**: Atom A has donated an electron to Atom B. As a result, Atom A becomes positively charged (A+) and Atom B becomes negatively charged (B-). The dashed red line indicates the attraction between these oppositely charged ions.
3. **Hydrogen Bonding**: The hydrogen atom (H) bonded to the oxygen atom (O) is attracted to another oxygen atom. The solid red line indicates the covalent bond within the molecule, while the dashed green line represents the weaker hydrogen bond.

_I hope this helps clarify_ the differences between covalent, ionic, and hydrogen bonds!
**BingChat:**

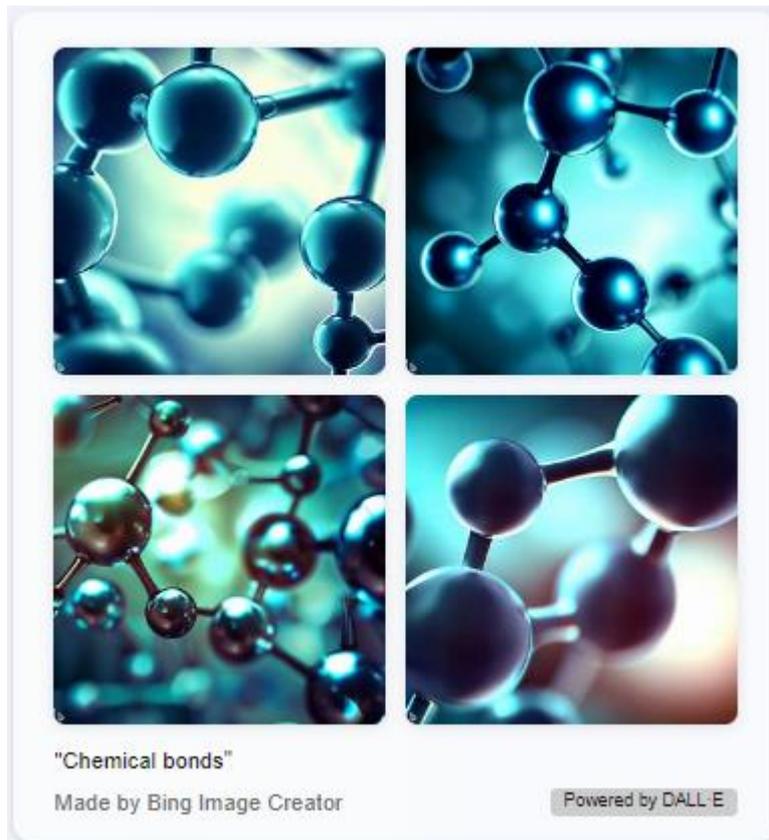

**Bard:** <u>Sure. Here is an illustration</u> explaining the different types of chemical bonds:
Ionic bonds

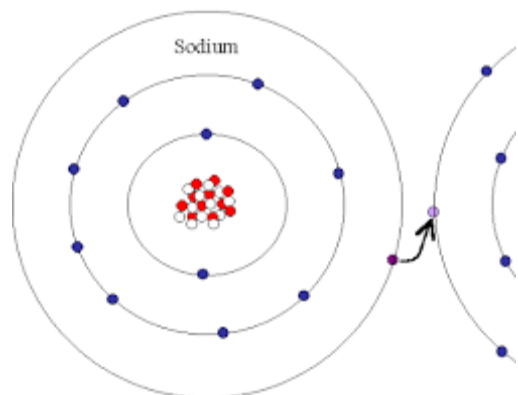

ionic bond between a sodium atom and a chlorine atom
In an ionic bond, one atom transfers one or more electrons to another atom. This results in two oppositely charged ions, which are attracted to each other by electrostatic forces.
In the image above, the sodium atom transfers one electron to the chlorine atom. The sodium atom becomes a positively charged ion (Na+), and the chlorine atom becomes a negatively charged ion (Cl-). The Na+ and Cl- ions are attracted to each other by electrostatic forces, and they form an ionic bond.
Covalent bonds

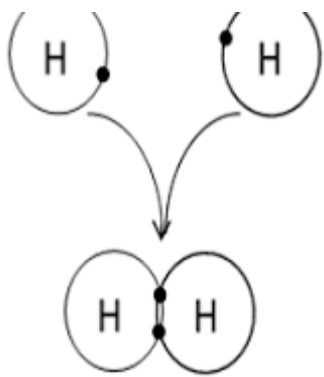

*covalent bond between two hydrogen atoms*
*In a covalent bond, two atoms share one or more pairs of electrons. Covalent bonds are typically formed between nonmetals.*
*In the image above, the two hydrogen atoms share a pair of electrons. The two hydrogen atoms are covalently bonded together.*
*Hydrogen bonds*

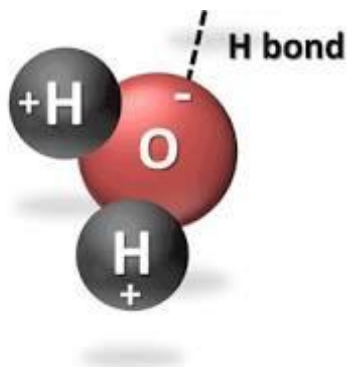

*hydrogen bond between two water molecules*
*A hydrogen bond is a special type of weak bond that forms between a hydrogen atom and a highly electronegative atom, such as oxygen or nitrogen. Hydrogen bonds are responsible for many of the unique properties of water, such as its high boiling point and surface tension.*
*In the image above, the hydrogen atom in one water molecule is attracted to the oxygen atom in another water molecule. This hydrogen bond is weaker than a covalent bond, but it is strong enough to hold the two water molecules together.*
*I hope this illustration helps you to understand the different types of chemical bonds. Please let me know if you have any other questions.*
**Claude:** *Here is an illustrated explanation* of the three main types of chemical bonds - ionic, covalent, and hydrogen bonds:

```
Ionic Bond:
+ Na + Cl -
The electron from sodium is completely
transferred to chlorine, forming ions that
attract electrostatically.

Covalent Bond:

H:H
The electrons are shared between the
hydrogen atoms, with both attracted to
the shared pair.

Hydrogen Bond:

Hδ+ Oδ-
The hydrogen side of the water molecule
has a slight positive charge from the
electronegative oxygen tugging electrons.
This attracts the negative end of nearby
water molecules.
```

*In this illustration:*
- *Ionic bonding is shown with the full electron transfer between the sodium and chlorine atoms, forming charged ions that attract.*
- *Covalent bonding is shown with the shared pair of electrons between two hydrogen atoms.*
- *Hydrogen bonding is shown with the polarity of the water molecule, leading to electrostatic attraction between molecules.*

<u>Let me know if</u> *a visual explanation like this helps differentiate the types of bonds!* <u>I can create more illustrated examples</u> *to demonstrate chemical concepts.* <u>Images and diagrams can sometimes bring these abstract ideas to life.</u>

ChatGPT starts with the usual **Reinforcement & Feedback** "Certainly!," followed by "Let's break down the explanation," which demonstrates **Instructional strategy**. After a detailed description of what the visualisation should comprise, repeatedly asking the student to "Imagine" something, it generated a quite crude "visual representation of the three types of bonds", finalising with a **Personalization & Personable Touch** "I hope this help clarify". **BingChat's** solely reliance on four high-quality, yet generic, pictures of chemical bonds not only did the images fail to employ effective **Instructional Strategies**, but they also did not promote genuine **Engagement &**

**Interactivity**, failing to deeply engage users in the topic. The content also lacked in **Use of Analogy & Comparative Illustrations**, as it missed drawing on parallels or real-world examples. Furthermore, the images also missed adding a **Personalization & Personable Touch**, making the learning experience feel less tailored and intimate. **Bard** used **Instructional strategy**, by providing a comprehensive explanation, accompanied by three specific images sourced from the internet, each distinctly depicting a type of bond, adding **Detail Depth & Content Recommendation**. Concluding its explanation, Bard added a touch of **Personalization & Personable Touch**, expressing, "I hope this illustration helps you" and prompting further engagement with "Please let me know..." **Claude**'s depiction was even more rudimentary, presenting bond details in textual format instead of offering a lucid visual illustration. Such a choice reflected a deficiency in **Instructional Strategies** and diminished **Engagement & Interactivity**, even though the final assertion "Images and diagrams can sometimes bring these abstract ideas to life." While the closing remarks of "Let me know" and "I can create more illustrated examples" hint at a desire for **Personalization & Personable Touch**, the lack of an actual tailored visual representation undermines this effort.

**Question about the essence of the type of chemical bonding**

*P1: so ionic bonding is when the atom gives the electron once and for all, in the covalent bond, it only lends and in the hydrogen bond it's like a bid, like the electron is with an atom but gives the atom a wink neighbor?*

*ChatGPT: <u>That's a creative and fun way to describe them, and it's largely on point! Let's expand on your analogies</u>:*

1. ***Ionic Bonding:***
   - Yes, it's like one atom <u>saying, "Here, take my electron. I don't need it."</u> As a result, one atom becomes positively charged (having given away its electron) and the other becomes negatively charged (having accepted the electron). These opposite charges attract each other, forming the bond.
2. ***Covalent Bonding:***
   - It's a mutual agreement between atoms: "<u>Let's share our electrons so we both benefit</u>." The electrons are not strictly lent or borrowed; instead, they're shared between the atoms, spending time around both nuclei.
3. ***Hydrogen Bonding:***
   - <u>Your analogy of a "wink" is a fun way to think of it</u>. In hydrogen bonding, the hydrogen atom is indeed bonded covalently to its original atom (like oxygen in water). But because hydrogen has a slight positive charge and nearby electronegative atoms (like another oxygen) have a slight negative charge, they're attracted to each other. It's not a full-on bond like in covalent or ionic bonding but rather a weaker attraction. <u>The "wink" analogy captures this idea of a subtle yet meaningful connection</u>.

*It's essential to remember that these analogies, while helpful, are simplifications. The actual forces and behaviors of atoms and molecules are based on deeper quantum-mechanical principles. But for getting a grasp on the concepts, these analogies work quite well!*
*Is there anything else about these bonds or any other related concept you'd like to delve into?*
**BingChat:** *That's a very creative way of describing the different types of bonds, but it's not very accurate. Let me try to correct some of the mistakes in your analogy.*
*Ionic bonding is not when the atom gives the electron once and for all, but rather when the atom transfers the electron permanently to another atom. This means that the atom that loses the electron becomes positively charged, and the atom that gains the electron becomes negatively charged. These opposite charges attract each other and form an ionic bond. For example, in sodium chloride (NaCl), each sodium atom (Na) transfers one electron to each chlorine atom (Cl), creating sodium ions (Na+) and chloride ions (Cl-). The Na+ and Cl- ions are held together by ionic bonds.*
*Covalent bonding is not when the atom only lends the electron, but rather when the atom shares the electron with another atom. This means that both atoms contribute one or more electrons to form a pair or a group of electrons that belong to both atoms. These shared electrons form a covalent bond that holds the atoms together. For example, in water (H2O), each hydrogen atom (H) shares one electron with the oxygen atom (O), forming a covalent bond. The O and H atoms are held together by covalent bonds.*
*Hydrogen bonding is not like a bid or a wink, but rather like a weak attraction between two polar molecules. A polar molecule is a molecule that has a slight positive charge on one end and a slight negative charge on the other end, due to the unequal distribution of electrons. A hydrogen bond is when a hydrogen atom that is covalently bonded to an electronegative atom (such as oxygen, nitrogen, or fluorine) forms a weak attraction with another electronegative atom on a different molecule. For example, in liquid water, each hydrogen atom on one water molecule can form a hydrogen bond with the oxygen atom on another water molecule.*
*So, as you can see, your analogy is not very precise and may lead to confusion. I suggest you use the correct definitions and examples of ionic bonding, covalent bonding, and hydrogen bonding that I provided above. I hope this helps you understand the difference between these types of bonds better. 😊*
**Bard:** *Your analogy is pretty good! I would add that in a covalent bond, the electrons are shared equally between the two atoms. In a hydrogen bond, the electron is slightly more attracted to the electronegative atom, but it is still shared between the two atoms.*
*Here is a more detailed explanation:*
*Ionic bonds are formed when one atom transfers one or more electrons to another atom. This results in two oppositely charged ions, which are attracted to each other by electrostatic forces.*
*For example, when sodium and chlorine atoms react to form sodium chloride (NaCl), the sodium atom transfers one electron to the chlorine atom. The sodium atom becomes a positively charged ion (Na+), and the chlorine atom becomes a negatively charged ion (Cl-). The Na+ and Cl- ions are attracted to each other by electrostatic forces, and they form an ionic bond.*
*Covalent bonds are formed when two atoms share one or more pairs of electrons. Covalent bonds are typically formed between nonmetals.*
*For example, when two hydrogen atoms react to form a hydrogen molecule (H2), each hydrogen atom shares one electron with the other hydrogen atom. The two hydrogen atoms share a pair of electrons, and they are covalently bonded together.*
*Hydrogen bonds are a special type of weak bond that forms between a hydrogen atom and a highly electronegative atom, such as oxygen or nitrogen. Hydrogen bonds are responsible for many of the unique properties of water, such as its high boiling point and surface tension.*

*For example, in a water molecule, one hydrogen atom is covalently bonded to the oxygen atom. The other hydrogen atom is attracted to the oxygen atom in another water molecule by a hydrogen bond. This hydrogen bond is weaker than a covalent bond, but it is strong enough to hold the two water molecules together.*
*<u>I hope this helps to clarify</u> the difference between ionic bonds, covalent bonds, and hydrogen bonds. <u>Please let me know</u> if you have any other questions.*
**Claude:** *<u>I love the "wink neighbor" analogy for hydrogen bonding! That's a very intuitive way to understand it.</u> To recap the key differences:*
- *Ionic bond: The electron is transferred permanently from one atom to the other. It's like giving away your Lego piece to someone else once and for all.*
- *Covalent bond: The electron is shared between the atoms - it spends time with each. It's like two friends sharing a Lego piece cooperatively.*
- *Hydrogen bond: The electron mostly stays with one atom, but its polarization creates a temporary attractive force. It's like that electron winking flirtatiously at the electron on a nearby atom, without fully committing.*

*<u>Your descriptions really capture the essence of each type of bond in a more fun and relatable way. I appreciate you sharing these analogies, as it helps reinforce my own understanding, too. Let me know if</u> there are any other aspects of chemical bonding you want me to try explaining conversationally like this!*

**ChatGPT** exemplifies **Personalization & Personable Touch** and the **Use of Analogy & Comparative Illustrations** by embracing the student's humorous and simplified analogy, agreeing that it captures the essence of each type of interaction with the remark, "Your analogy of a "wink" is a fun way to think of it". Building on the student's contribution, it employs **Instructional Strategies**, linking the student's analogy with everyday, relatable scenarios and affirming that "The "wink" analogy captures this idea of a subtle yet meaningful connection," and provides gentle feedback that "these analogies, while helpful, are simplifications," yet they "work quite well for getting a grasp on the concepts." Finally, to foster continued engagement, it prompts, "Is there anything else…," showcasing **Engagement & Interactivity**. **BingChat** exhibits a strong inclination towards **Reinforcement & Feedback**, directly dismissing the student's analogy as "not very accurate" and stating that "your analogy is not very precise and may lead to confusion." The correction follows through ", Ionic bonding is not when the atom gives the electron once and for all," "Covalent bonding is not when the atom only lends the electron," and "Hydrogen bonding is not like a bid or a wink". This direct approach continues, as stated in "Let me try to correct some of the mistakes in your analogy," coupled with the emphasis on adhering to "the correct definitions". Amidst this feedback-heavy approach, its concluding note of "I hope this helps" and the appended emoticon subtly gesture towards **Personalization & Personable Touch** might be overshadowed by the preceding corrective tone, risking overshadowing the student's initial enthusiasm and being perceived as somewhat dismissive. **Bard** commendably recognises the student's analogy, stating, "Your analogy is pretty good!" a nod to both **Personalization & Personable Touch** and the **Use of Analogy & Comparative Illustrations**. Diving deeper, it delves into **Detail Depth & Content Recommendation** by proffering a "more detailed explanation" of bonding types, replete with examples, merging the acknowledgement of the student's perspective with a clear, factual elucidation and elucidating on the formation of the bonding and consequential properties, accommodating those eager for comprehensive insights. Concluding its exposition with "I hope this helps to clarify" and "Please let me know," it seamlessly integrates **Personalization & Personable Touch** and **Engagement & Interactivity** into its response. **Claude**, in a similar vein, leans into **Personalization & Personable Touch** and the **Use of Analogy & Comparative Illustrations**, taking a conversational stance and appreciating the student's analogies with "I love the 'wink neighbor' analogy" and "That's a very intuitive way to understand it." It then leans to **Instructional Strategies**, by presenting a short explanation and extensively shows again appreciation for the student's approach with "Your descriptions really capture the essence of each type of bond in a more fun and relatable way. I appreciate you sharing these analogies, as it helps reinforce my own understanding too," which also indicates **Meta-discussion & Perspective Framing**, as it aids mutual understanding. It finalises with a touch of **Personalization & Personable Touch,** prompting further engagement with "Let me know..."

**Further analysis**

This study's findings diverge from those of Gregorcic and Pendrill (2023), who contended that despite rectifying errors and contradictions via Socratic dialogue, ChatGPT did not meet the necessary criteria to function as a physics tutor. This disparity could stem from using an older, less advanced iteration of ChatGPT, specifically the GPT-3 model introduced by OpenAI in 2020 (Brown et al., 2020).

Our research indicates that the current versions of ChatGPT, Bing Chat, Bard, and Claude can act as effective objects-to-think-with, as per Papert's concept (1980, p. 11), cultivating an engaging, interactive, and inclusive learning environment. These AI tools can support creativity, collaboration, exploration, critical thinking, and problem-solving, potentially deepening understanding of the subject matter. This relates to the comparison between ChatGPT, Bard, and Bing Chat in the study of Liu et al. (2023).

The overall success of this educational journey, with GenAIbots serving as agents to think with, highlights a few effective teaching strategies that can facilitate a profound comprehension of the involved concepts. A significant part of the process involved positive reinforcement, as GenAIbots acknowledged students' observations as enjoyable, validating their experiences and stimulating their curiosity. Furthermore, following Gregorcic and Pendrill (2023), we adopted what could be seen as a Socratic-like dialogue strategy, prompting students to think critically and actively engage in their learning rather than merely providing them with answers. GenAIbots also capitalised on the student's existing knowledge by inquiring about their understanding of topics like weight and gravity, thereby establishing a meaningful bridge between new and previously learned information.

Building upon Bitzenbauer's preliminary study (2023) that highlighted the potential of expansive language model-based chatbots, including ChatGPT, to enrich learning experiences, and Adiguzel et al.'s (2023) assertion of ChatGPT's transformative potential in education via personalised instruction and real-time feedback, our research emphasises the benefits of AI integration in learning environments. AI-driven chatbots like ChatGPT, Bing Chat, Bard, Claude, and other generative models might serve as advanced agents for addressing challenging learning scenarios, expanding students' knowledge incrementally, and providing relevant explanations, examples, and analogies to support their learning. This study highlights the potential of AI-driven tools in promoting active learning, personalised instruction, and conceptual understanding in Chemistry. It underscores the value of AI in enhancing educational experiences, concluding that GenAIbots can serve as innovative platforms for exploring challenging educational scenarios.

Nevertheless, our analysis reveals that ChatGPT and Claude consistently provided more comprehensive, detailed, and accurate answers than Bing Chat, which tended to give shorter and less informative responses. ChatGPT-4 was able to adapt responses to students' questions, address misconceptions, and offer customised explanations. ChatGPT was also more adept at addressing the nuances and subtleties of the questions, recognising the context, and engaging with the user's unique perspective or interpretation. On the other hand, Bing Chat often failed to address these nuances and provided less contextual information. Overall, ChatGPT and Claude demonstrated a superior understanding of the subject matter and a more remarkable ability to convey complex scientific concepts in an accessible manner.

The study also highlighted the need for comprehensive educator training before integrating these tools into classrooms. Adequate training empowers teachers to guide students effectively in articulating their thoughts and constructing solid arguments. This preparation would allow educators to assess GenAIbots' performance when faced with compelling counterexamples, fostering productive discussions among colleagues.

Finally, it is essential to mention that the results of this experience revealed significant variance in the performances of ChatGPT, Bing Chat, Bard, and Claude. This result agrees with Franciscu (2023), in which ChatGPT, a cutting-edge natural language generation model, excels at producing human-like responses suitable for chatbots. Based on the detailed analysis above, despite some fundamental similarities, ChatGPT notably outperformed BingChat, with Bard and Claude closely shadowing ChatGPT's performance and displaying comparable proficiency levels between them, as described below and summarised in Table 2.

1. **ChatGPT** stands out for its consistent **Personalization & Personable Touch** through positive reinforcement and feedback mechanisms. It employs effective **Instructional Strategies**, often breaking down explanations systematically, and frequently engages users with **Engagement & Interactivity** by soliciting feedback. With a flair for the **Use of Analogy & Comparative Illustrations**, ChatGPT translates abstract concepts into relatable scenarios. Its **Detail Depth & Content Recommendation** is tailored and comprehensive, and it occasionally ventures into **Meta-discussion & Perspective Framing** to provide broader contexts.
2. **Claude**'s strengths lie predominantly in **Personalization & Personable Touch** and **Engagement & Interactivity**, making learning feel like an interactive dialogue. Offering straightforward explanations, it leans heavily on the **Use of Analogy & Comparative Illustrations**, especially when building upon user insights. While its **Instructional Strategies** provide clarity, the depth can vary, catering to concise and elaborate queries. Claude's penchant for **Meta-discussion & Perspective Framing** emphasises the broader significance of concepts, enhancing the overall learning experience.
3. **Bard**'s responses resemble a classic **Instructional Strategies** approach, akin to a conventional classroom style. Focusing on factual and structured presentations, it often employs the **Use of Analogy & Comparative Illustrations** to elucidate concepts. **Personalization & Personable Touch** is evident, especially when recognising user inputs and providing deeper insights, fostering **Engagement & Interactivity**. Bard occasionally integrates **Meta-discussion & Perspective Framing** to highlight broader applications.
4. **BingChat** emphasises **Detail Depth & Content Recommendation** by often directing users to external resources. Adopting a more correctional **Reinforcement & Feedback** approach, it points out inaccuracies in user inputs, which can sometimes overshadow its **Personalization & Personable Touch**. While BingChat provides structured

**Instructional Strategies**, its reliance on external links might diminish direct **Engagement & Interactivity**. Its **Use of Analogy & Comparative Illustrations** is present, albeit less pronounced.

This ranking is based on the analysis presented earlier. It provides a comparative grade for each AI across different categories, but the specific grading might be subjective and could vary with different questions, contexts or even across sessions. The preference might also be subjective depending on individual learning styles and needs.

**Table 2** Comparative Analysis of ChatGPT, Bing Chat, Bard, and Claude on this Experience

| Categories | ChatGPT | BingChat | Bard | Claude |
|---|---|---|---|---|
| **Personalization & Personable Touch** | Very High | Fair | Good | High |
| **Instructional Strategies** | Very High | Good | High | Good |
| **Engagement & Interactivity** | High | Fair | Fair | Good |
| **Use of Analogy & Comparative Illustrations** | High | Fair | Good | High |
| **Detail Depth & Content Recommendation** | High | Good | High | Good |
| **Reinforcement & Feedback** | High | Fair | Good | Good |
| **Meta-discussion & Perspective Framing** | Good | Fair | High | Good |

Using GenAIbots as educational tools, teachers might create more engaging and personalised learning experiences. This approach could potentially support a deeper understanding of complex concepts in Chemistry and perhaps other subjects.

**Conclusions**

In our study evaluating the proficiency of four AI language models—ChatGPT, Bing Chat, Bard, and Claude—in elucidating chemistry concepts, ChatGPT stood out, decisively surpassing Bing Chat's performance. Bard and Claude trailed closely, with all three showcasing a more in-depth, precise, and nuanced understanding, underscoring ChatGPT's adeptness at contextual comprehension.

Our findings highlight the robustness of GenAIbots like ChatGPT, Bing Chat, Bard, and Claude as agents-to-think-with in Chemistry education. These AI models seem suitable to offer interactivity, customisation, and multidisciplinary knowledge and promote critical thinking, problem-solving, creativity, and collaboration. They facilitate dynamic conversations and tailored content in Chemistry and potentially other domains.

Engaging with GenAIbots might enhance critical thinking, problem-solving, and creativity, fostering collaborative discussions and projects in group settings and deepening understanding of the topics. By integrating GenAIbots into Chemistry education, we're aligning with the principles of Papert's Constructionism. This theory posits that learning is most effective when students actively construct knowledge by creating tangible representations in meaningful contexts. This approach encourages students to engage deeply, fostering their problem-solving, critical thinking, and creative skills.

**Limitations of the study and possible future studies**

Despite its inherent limitations, including its single case design and the potential for bias, the exploratory depth of the study uncovered hidden potential within these systems, even amid serious concerns about generalizability.

Future research could include:
- Refining the crafting of prompts.
- Exploring new features of these and other GenAIbots being introduced with increasing frequency.
- Establishing standardised benchmarks to evaluate and compare chatbots and AI systems' performance, accuracy, and reliability.
- Conducting long-term studies to observe the evolution of chatbots' capabilities and their impact on user interactions over time.
- Conducting research with real students in classroom settings and beyond to assess these AI systems' practical educational applications and challenges.
- Investigating chatbots' learning and adaptation capabilities to individual user needs and preferences over time.
- Researching the integration of multimodal inputs (e.g., text, voice, image) to enhance chatbot capabilities and user interaction experiences.

When implementing GenAIbots in Chemistry learning, it is crucial to evaluate benefits and drawbacks judiciously, ensuring accurate information delivery and considering the implications of reduced human interaction. These concerns may be alleviated by integrating GenAIbots with other educational tools or activities promoting collaborative dialogue among learners.

**Author Contributions**

The author confirms sole responsibility for the following: study conception and design, data collection, analysis and interpretation of results, and manuscript preparation.

**Conflicts of interest**

There are no conflicts to declare.

**Acknowledgements**

The authors warmly acknowledge Melanie Swan for her invaluable suggestion, which led to the transition from using the term 'objects-to-think-with' to 'agents-to-think-with'.